\documentclass[a4paper,fleqn]{cas-dc}
\usepackage[numbers]{natbib}
\usepackage{amsfonts}
\usepackage{algorithmic}
\usepackage{amsmath} % matrices
\usepackage{amssymb} % mathbb{}
\usepackage{caption}
\usepackage{subcaption}
\usepackage{etoolbox}
\usepackage{forest}
\usepackage{lingmacros}
\usepackage{textcomp}
\usepackage{tree-dvips}
\usepackage{tikz}
\usepackage{tikz-cd}
\usepackage[arrowdel]{physics}
\usepackage{graphicx}
\usepackage{wrapfig}
\usepackage{listings}
\usepackage{pgfplots, pgfplotstable}
\usepackage{diagbox} % diagonal line in cell
\usepackage[usestackEOL]{stackengine}
\usepackage{makecell}
\usepackage{mathrsfs}
\usepackage{moresize}
\usepackage{multirow}
\usepackage{multicol}
\usepackage[numbers]{natbib}
\usepackage[T1]{fontenc}
\usepackage{xcolor}
\allowdisplaybreaks[1]
\definecolor{orchid}{rgb}{0.7, 0.4, 1.1}
\definecolor{comment_color}{rgb}{0, 0.5, 0}
\definecolor{keyword_color}{rgb}{0.3, 0, 0.6}
\definecolor{string_color}{rgb}{0.5, 0, 0.1}

\begin{document}
\shorttitle{Quantum optoelectronics in semiconductor solar cell materials and devices}

\shortauthors{Xi Liu$^a$, Wenxi Fang$^b$, Ken Perlin$^c$}

\title{{\Large Quantum optoelectronics in semiconductor solar cell materials and devices}}
% \author{Xi Liu}
% \email{xl3467@columbia.edu}
% \affiliation{Columbia University, Electrical\,Engineering, 10027, New\,York, United\,States}
\author[1]{\color{black}Xi Liu}
\author[2]{\color{black}Wenxi Fang}

\address{$^a$xl3467@columbia.edu, Columbia University; $^b$u3013972@connect.hku.hk, Inner Mongolia University of Science and Technology}
\begin{abstract}
% Quantum optoelectronics explores the interaction between light and matter at the particle, atom, and molecular level, enabling the development of advanced technologies in energy storage and computing. This paper delves into fundamental concepts such as cavity quantum electrodynamics (CQED), Fabry Perot resonators, photovoltaic materials, and their integration into optical devices for solar cells. This paper analyzes the performance characteristics of key semiconductor solar cell materials, perovskites, transition metal dichalcogenides (TMDs), cadmium telluride (CdTe), and silicon, and examines their interrelation with foundational concepts in quantum optoelectronics and optomechanics. By analyzing recent advancements and integrating principles from quantum photonics, we highlight how these materials benefit from and contribute to the development of next generation photovoltaic technologies. We examine the underlying principles, recent advancements, and potential applications that position quantum optoelectronics at the forefront of next generation photonic technologies.
% Quantum optoelectronics is an interdisciplinary field that combines principles from quantum mechanics, photonics, and materials science to enhance light-matter interactions for energy applications.
We analyze the integration of quantum optical phenomena, such as cavity quantum electrodynamics (CQED), Fabry Perot resonances, and strong light-matter coupling, into the design and engineering of next generation photovoltaic systems. We examine how these phenomena can be harnessed through photonic structures including optical cavities, plasmonic materials, and metasurfaces to improve light trapping, absorption, and carrier dynamics in future solar cell devices. Specific focus is given to semiconductor materials such as perovskites, organics, transition metal dichalcogenides (TMD), cadmium telluride (CdTe), and silicon. For perovskite solar cells, we analyze device architectures, interfacial engineering with hyperbranched polymers, and additive optimization using molecular dopants and nanosheets to enhance film morphology and stability. We further examine laser-based metrology for thin-film characterization and coherent spectroscopy techniques involving frequency combs and high-harmonic generation. The paper also shows how machine learning (ML), combined with density functional theory (DFT), accelerates material screening and performance prediction for next-generation solar cell absorbers. These developments demonstrate how quantum optoelectronic design principles are transforming photovoltaic research and enabling higher efficiency, stability, and functionality in solar energy devices.
\end{abstract}
\maketitle
\section{Introduction}

The field of quantum optoelectronics merges quantum mechanics with optical engineering to manipulate and control light-matter interactions at the quantum scale. This interdisciplinary domain has led to significant progress in developing devices like single-photon sources, quantum switches, and high-precision sensors. Central to these advancements are structures such as optical cavities and resonators, which enhance and control quantum interactions.

Cavity quantum electrodynamics (CQED) studies the interaction between quantum emitters (like atoms or quantum dots) and the electromagnetic field within a confined cavity. The confinement alters the density of photonic states, leading to phenomena such as vacuum Rabi oscillations and the Purcell effect.
% The 2012 Nobel Prize in Physics was awarded to Serge Haroche and David Wineland for their pioneering work in CQED, demonstrating control over individual quantum systems and laying the groundwork for quantum computing and metrology.
The Jaynes-Cummings model mathematically describes the interaction between a two-level atom and a single mode of the quantized electromagnetic field, predicting energy level splitting and coherent oscillations between the atom and the field. These interactions are fundamental for quantum information processing and communication.

The integration of optical cavities into solar cell designs has become a promising strategy to enhance light-matter interaction and improve power conversion efficiency \cite{steinfeld_optimum_1993}. Optical cavities, such as Fabry Perot resonators, can be engineered to increase the dwell time of photons within the active layer of a solar cell. By tailoring the cavity dimensions and the refractive index contrast, constructive interference patterns can be created to trap incident sunlight at specific resonant wavelengths, especially those where the absorber material has low absorption coefficients \cite{prakash_2009}. This optical path length enhancement boosts the generation of electron-hole pairs without increasing the material thickness, which is especially advantageous for thin-film and organic photovoltaics.

Laser processing and characterization serve multiple roles in the development of high-efficiency solar cells. Ultrafast laser pulses can be used for surface texturing to reduce reflection losses or to create photonic crystal patterns for light trapping. Furthermore, lasers are instrumental in photoluminescence and time-resolved spectroscopy to study carrier lifetimes and defect states in solar materials. In the emerging field of upconversion solar cells, coherent laser sources are used to test nonlinear optical responses where sub-bandgap photons are converted to above-bandgap energies through multiphoton processes, enabling broader spectrum utilization.

Photonic components, including waveguides, metamaterials, and plasmonic structures, play a crucial role in directing and manipulating light within solar cells. Nanostructured photonic crystals can create forbidden energy bands for photons, leading to slow-light effects and enhanced absorption. Plasmonic nanoparticles can concentrate the electromagnetic field at the nanoscale, increasing the effective absorption cross-section of nearby semiconductors. Hybrid photonic-plasmonic systems are actively studied for their ability to synergize both broadband light trapping and hot-electron injection, potentially boosting the internal quantum efficiency of photovoltaic devices.

In solar cell applications, incorporating cavity structures can enhance light absorption and emission properties \cite{weinstein_2014}. For instance, modeling PSCs as Fabry Perot resonators has been used to calculate ideal power conversion efficiencies, taking into account interference effects and optimizing the optical flux within the absorber layer \cite{weinstein_absorption_2015}. Optomechanics involves the interaction between optical fields and mechanical vibrations. In the context of solar cells, integrating optomechanical elements can lead to dynamic control of light absorption and scattering \cite{betancur_2010}, potentially enhancing device performance. While still an emerging area, the principles of optomechanics offer promising avenues for future photovoltaic technologies \cite{harris_thermal_1985}.

Machine learning offers powerful tools for material discovery, performance prediction, and device optimization in solar cell research \cite{tao_2021}. Supervised learning algorithms can predict photovoltaic properties such as bandgap, carrier mobility, and stability based on chemical composition and structural features. Unsupervised clustering can identify new material classes with promising properties from large datasets \cite{chen_peroskite_2023}. In optical design, neural networks can rapidly simulate complex photonic structures or inverse-design metasurfaces for optimal light management \cite{zhi_2023_screen}. Reinforcement learning is also being explored for autonomous experimentation, where robotic systems iteratively synthesize and test materials under the guidance of a learning algorithm, accelerating the search for next-generation solar materials \cite{zhang_2022_add}.

\section{Next generation quantum optoelectronic devices for solar cells}
Solar energy represents one of the most sustainable and cleanest alternatives to fossil fuel-based electricity generation, playing a vital role in reducing global pollution, energy waste, and environmental degradation.
Fossil fuels, such as oil, natural gas, and coal, are non-renewable energy sources with limited availability, and continued high rates of consumption are accelerating their depletion. According to current projections, global oil supplies may be exhausted within the next 50 years. Natural gas reserves are expected to endure for roughly 90 to 120 years. Coal could last around 130 years if usage patterns remain unchanged.
Unlike coal, oil, or natural gas combustion, which emits large quantities of carbon dioxide and toxic pollutants, solar photovoltaics directly convert sunlight into electricity without producing greenhouse gases or hazardous byproducts. This not only mitigates air pollution and health hazards but also contributes to combating climate change. Furthermore, solar power enables decentralized energy generation, reducing transmission losses that occur in long-distance power grids. The versatility of solar cells extends across a broad range of applications: from powering electronic devices, to enabling solar-powered transportation systems. In aerospace, solar cells are indispensable for satellites, spacecraft, and space stations, where solar energy is often the only viable power source. Solar installations serve industrial facilities, commercial centers, and residential buildings, reducing reliance on the central grid and lowering electricity costs. By integrating solar energy across these domains, we not only enhance energy efficiency but also advance toward a more sustainable, cleaner future.

\subsection{Fabry Perot resonators}

A Fabry Perot resonator consists of two parallel, partially reflective mirrors that create a standing wave cavity for light. Only specific wavelengths that satisfy the resonance condition constructively interfere and are transmitted, making these resonators essential for applications requiring precise wavelength selection. The finesse of a Fabry Perot cavity, defined by the reflectivity of the mirrors and the separation distance, determines the sharpness of the resonance peaks. High-finesse cavities are crucial for enhancing light-matter interactions in CQED experiments. Recent advancements include the development of nanofiber-based Fabry Perot microresonators, which integrate fiber Bragg gratings to form compact, high-finesse cavities suitable for nonlinear optics and CQED applications.

Multiple exciton generation (MEG) is a process where a single high-energy photon generates multiple electron-hole pairs, potentially increasing the photocurrent beyond the Shockley-Queisser limit. This phenomenon has been observed in materials like PbSe quantum dots and is being explored in various semiconductor nanostructures, including perovskites and transition metal dichalcogenides (TMD) \cite{haiyan_zheng_fabry_2023}. Harnessing MEG effectively could lead to significant enhancements in solar cell efficiencies.

\begin{figure}
\includegraphics[width = 0.45\textwidth, height = 0.3\textwidth]{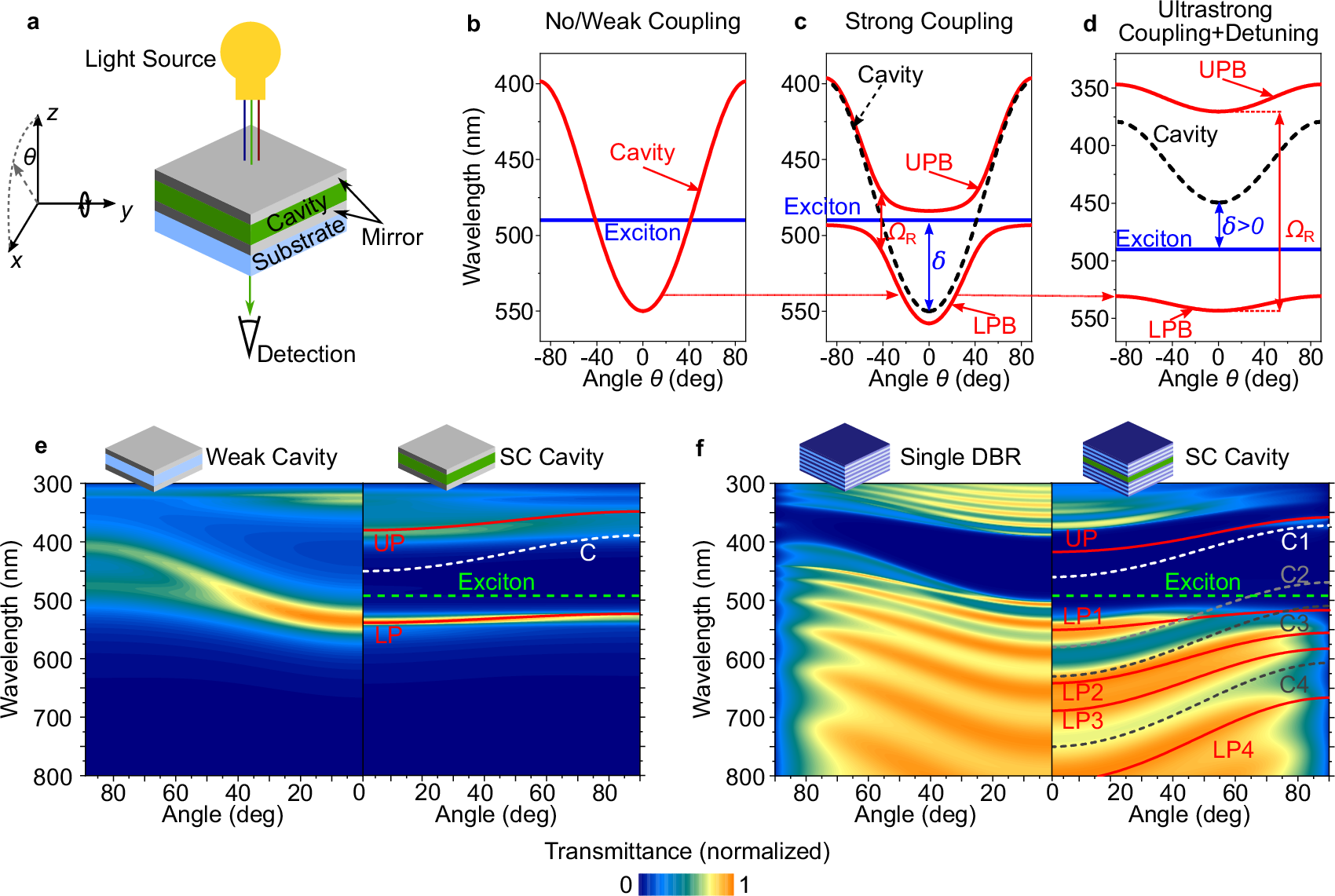}
\caption{polariton dispersion management. (a). cavity-based polariton transmission filter. (b-d). angular dispersion simulation using transfer matrix \cite{mischok_angular_2024}}
\label{angular dispersion limit in thin film optics, mischok}
\end{figure}

Figure \ref{angular dispersion limit in thin film optics, mischok} illustrates the impact of varying light-matter coupling strengths on the angular dispersion of optical microcavities, highlighting how ultra-strong coupling can effectively manage dispersion.
In the weak coupling regime, a Fabry Perot microcavity filled with a transparent material like SiO$_2$ exhibits a parabolic dispersion relation. This means that the resonance wavelength shifts significantly with changes in the angle of incidence, leading to angular dispersion.
Introducing a material with a strong excitonic resonance, such as the organic dye C545T, into the cavity leads to strong light-matter coupling. This results in the formation of two new hybrid modes: the lower polariton branch (LPB) and the upper polariton branch (UPB). The LPB is red-shifted relative to both the photon and exciton resonances and exhibits a flatter dispersion at larger angles, while the UPB shows the opposite behavior.
% By tuning the microcavity to a relatively large coupling strength and a moderate positive detuning (the energy difference between the bare photon and exciton resonances), the LPB becomes nearly flat across a range of angles. This configuration minimizes angular dispersion and maintains high transmission. Further increasing the positive detuning can reduce peak transmission, while negative detuning reintroduces angular dispersion by making the LPB more photonic in character.
Utilizing ultra-strong coupling not only increases the separation between the LPB and UPB, allowing for selective filtering, but also separates the LPB from the bare exciton absorption. This separation reduces parasitic absorption from uncoupled excitons and maintains a flat dispersion profile.
Transfer matrix simulations comparing cavities with transparent SiO$_2$ and those with C545T demonstrate that ultra-strong coupling in the latter leads to a significant Rabi splitting and a near-complete flattening of the dispersion. This effect is not limited to metallic cavities; distributed Bragg reflector (DBR) cavities with alternating high and low refractive index layers can also exhibit strong coupling with excitonic resonances, leading to multiple Bragg-polariton branches with reduced angular dispersion.
These findings suggest that ultra-strong coupling in microcavities offers a powerful approach to managing dispersion in optical devices, enabling the design of filters and coatings with minimal angular dependence and high transmission efficiency.

\subsection{Vertical cavity surface emitting lasers (VCSEL) and extreme ultraviolet laser spectroscopy}
VCSELs are semiconductor lasers with emission perpendicular to the wafer surface, utilizing distributed Bragg reflectors (DBRs) as mirrors. Their compact size, low threshold currents, and ability to form two-dimensional arrays make them suitable for high-speed data communication and sensing applications. In quantum optoelectronics, VCSELs can be engineered to emit single photons or entangled photon pairs, essential for quantum communication protocols.
% The generation of coherent extreme ultraviolet (EUV) radiation is pivotal for advancing high-resolution spectroscopy, particularly in probing narrow nuclear transitions such as the $^{229\text{m}}$Th isomeric state. A Yb-fiber frequency comb can be employed to generate EUV light through cavity-enhanced seventh-harmonic generation \cite{zhang_noncollinear_2020}.
The pursuit of high-precision spectroscopy in the vacuum ultraviolet (VUV) and EUV spectral regions necessitates light sources that combine coherence, tunability, and high spectral resolution \cite{porat_2018}. Traditional laser sources often fall short in this regard due to limitations in wavelength coverage and coherence properties. High-harmonic generation (HHG) has emerged as a viable technique to bridge this gap, enabling the upconversion of near-infrared (NIR) laser pulses to the EUV regime.

\subsubsection{Theory of laser frequency comb}
An optical fiber laser frequency comb is a laser source whose spectrum consists of discrete, equally spaced frequencies. These frequencies can be expressed as
\[
f_n = f_{\text{CEO}} + n f_{\text{rep}}
\]
where \( f_n \) is the frequency of the \( n \)-th comb tooth \cite{adler_2004}, \( f_{\text{rep}} \) is the pulse repetition rate (i.e., the spacing between the teeth), \( f_{\text{CEO}} \) is the carrier-envelope offset frequency, and \( n \in \mathbb{Z} \) is an integer typically on the order of \(10^5\) to \(10^6\) \cite{bao_2019}.

In a mode-locked fiber laser, many longitudinal modes of the laser cavity are phase-locked, leading to the formation of a pulse train. The repetition rate \( f_{\text{rep}} \) is determined by the round-trip time of the pulse in the cavity
\[
f_{\text{rep}} = \frac{1}{T_R} = \frac{c}{n_{\text{eff}} L}
\]
where \( T_R \) is the cavity round-trip time, \( c \) is the speed of light in vacuum, \( n_{\text{eff}} \) is the effective refractive index of the fiber, and \( L \) is the optical path length of the cavity.

The carrier-envelope offset frequency \( f_{\text{CEO}} \) arises due to the difference between the group and phase velocities in the fiber \cite{weiner_2017}. The electric field of the pulse train in time domain can be written as
\[
E(t) = \sum_n A(t - n T_R) \cos(\omega_0 (t - n T_R) + \phi_n)
\]
where \( A(t) \) is the pulse envelope, \( \omega_0 \) is the carrier angular frequency, and \( \phi_n \) is the carrier-envelope phase shift \cite{tschernig_2024}. Due to dispersion, the carrier slips relative to the envelope with each round trip, resulting in:
\[
f_{\text{CEO}} = \frac{\Delta \phi_{\text{CE}}}{2\pi} f_{\text{rep}}
\]
where \( \Delta \phi_{\text{CE}} \) is the carrier-envelope phase shift per round trip.

The Fourier transform of this pulse train reveals its frequency-domain structure. A periodic train of pulses spaced by \( T_R \) has a frequency spectrum with lines spaced by \( f_{\text{rep}} \), shifted by \( f_{\text{CEO}} \). Thus, the frequency comb is composed of the set \( \{ f_{\text{CEO}} + n f_{\text{rep}} \} \).

To fully stabilize the comb, both \( f_{\text{rep}} \) and \( f_{\text{CEO}} \) must be measured and controlled. One common technique is \( f - 2f \) self-referencing. This involves generating the second harmonic of a lower-frequency comb tooth \( f_n \), and beating it against a higher-frequency tooth \( f_{2n} \):
\[
2 f_n = 2(f_{\text{CEO}} + n f_{\text{rep}}), \quad f_{2n} = f_{\text{CEO}} + 2n f_{\text{rep}}
\]
which gives a beat frequency of
\[
f_b = 2 f_n - f_{2n} = f_{\text{CEO}}
\]
This beat frequency can be used in a phase-locked loop to stabilize the offset frequency.

The experiment involves directing a train of EUV pulses from a frequency comb at trapped Mg-like Ar$^{6+}$ ions. Initially in the $^1S_0$ ground state, the ions are excited by the pulse train to a metastable state such as $^3P_1$ or $^3P_2$ \cite{nauta_2017}. Because these excited states have long lifetimes, they can coherently interact with many pulses before decaying, providing a means to probe the temporal coherence of the pulse train. If the coherence time is short, phase fluctuations between the pulses degrade the ion’s excitation over time \cite{fischer_2019}. To predict the behavior of the ions under excitation, multi-configuration Dirac-Hartree-Fock (MCDHF) theory was used to compute energy levels and transition lifetimes \cite{gustafsson_2017}. These are compared with experimental values from the NIST database. For example, the $^3P_1$ and $^3P_2$ states in Ar$^{6+}$ have lifetimes of approximately $1.3\ \mu$s and $5.6$ s, respectively, and transition energies of $14.12248(24)$ eV and $14.33133(25)$ eV.

The Dirac-Hartree-Fock (DHF) theory is a relativistic version of the Hartree-Fock (HF) method. It is used to describe the electronic structure of atoms and ions, especially in cases where relativistic effects such as spin-orbit coupling become significant \cite{sun_dirac_2021}. This is especially important for highly charged ions and heavy atoms. In DHF theory, each electron is described by a four-component Dirac spinor
\[
\psi_i(\vb{r}) =
\begin{pmatrix}
\varphi_i(\vb{r}) \\
\chi_i(\vb{r})
\end{pmatrix}
\]
where $\varphi_i$ and $\chi_i$ are the large and small components of the spinor, respectively. The full wavefunction of the multi-electron atom is approximated by a single Slater determinant composed of these spinors. The one-electron Dirac equation in the presence of the nuclear potential $V_N(\vb{r})$ is
\[
\left(c\boldsymbol{\alpha} \cdot \vb{p} + \beta m c^2 + V_N(\vb{r})\right)\psi_i(\vb{r}) = \epsilon_i \psi_i(\vb{r})
\]
$\boldsymbol{\alpha}$ and $\beta$ are the Dirac matrices, $c$ is the speed of light, $m$ is the electron mass, $\vb{p} = -i\hbar \nabla$ is the momentum operator, and $\epsilon_i$ is the single-electron energy eigenvalue. To account for electron-electron interactions, DHF uses the mean-field approximation. The full Dirac-Coulomb Hamiltonian for an $N$-electron atom is
\[
H = \sum_{i=1}^N \left(c \boldsymbol{\alpha} \cdot \vb{p}_i + \beta m c^2 + V_N(\vb{r}_i) \right) + \sum_{i<j} \frac{1}{|\vb{r}_i - \vb{r}_j|}
\]

In the Hartree-Fock approximation, the many-body problem is reduced to a set of coupled integro-differential equations for the single-particle orbitals. These equations for each orbital $\psi_i$ are:
\[
\left(h_D + V_{\text{HF}} \right) \psi_i = \epsilon_i \psi_i
\]
where $h_D$ is the Dirac operator for a single particle and $V_{\text{HF}}$ includes the direct (Coulomb) and exchange interactions arising from all other electrons. The mean-field potential acting on $\psi_i$ is
{\small
\begin{align*}
&V_{\text{HF}} \psi_i(\vb{r})\\
&= \sum_{j=1}^N \left(\int d^3 r'\frac{\psi_j^\dagger(\vb{r}')\psi_j(\vb{r}')\psi_i(\vb{r})}{|\vb{r} - \vb{r}'|} - \int d^3r'\frac{\psi_j^\dagger(\vb{r}')\psi_i(\vb{r}')\psi_j(\vb{r})}{|\vb{r} - \vb{r}'|}\right)
\end{align*}}
The first term is the direct (Coulomb) interaction and the second is the exchange term, which arises due to the antisymmetry of the total wavefunction.

Because the equations are nonlinear (each orbital depends on all the others), they are solved iteratively using a self-consistent field (SCF) method. One starts with an initial guess for the orbitals, computes the mean field, solves the Dirac equation to update the orbitals, and repeats until convergence. To include correlation and fine-structure effects beyond DHF, multiconfiguration Dirac-Hartree-Fock (MCDHF) is often used. In MCDHF, the atomic state is expressed as a linear combination of configuration state functions (CSF)
\[
\Psi = \sum_k c_k \Phi_k
\]
Each CSF $\Phi_k$ is constructed from a set of single-electron orbitals, and the coefficients $c_k$ are found by diagonalizing the Dirac-Coulomb Hamiltonian in the basis of these CSFs. This allows for configuration interaction and captures correlation effects more accurately.

In the time domain, the frequency comb electric field is modeled as a train of pulses
\[
E(t) = E_p \sum_j f(t - jT_r)\cos(\omega_0 t + \phi(t))
\]
where $E_p$ is the peak electric field, $f(t - jT_r)$ is the envelope of the $j$-th pulse spaced by a repetition time $T_r$, $\omega_0$ is the carrier frequency, and $\phi(t)$ represents the phase noise. In an ideal frequency comb, these parameters are perfectly stable, resulting in infinite coherence. However, real combs exhibit fluctuations in $\phi(t)$ and $T_r$, causing each frequency component (comb tooth) to broaden.

The phase noise $\phi(t)$ is modeled as a random walk process:
\[
\phi(t) = \int_0^t s(t') dt'
\]
where $s(t)$ is a Gaussian white noise process with autocorrelation $\langle s(t)s(t') \rangle = \sigma^2 \delta(t - t')$. This yields a coherence time $\tau_c = \frac{1}{2\pi \sigma^2}$, and results in a frequency tooth linewidth (FWHM) of $\sigma^2$.

The interaction between the EUV pulse train and the ion is modeled using Bloch equations in the rotating-wave approximation \cite{ziolkowski_1995}. Assuming a two-level system consisting of the ground state and one excited state, the density matrix elements evolve according to
\begin{align*}
&\dot{\rho}_{ee} = -\mathrm{Im}[\mu E(t) \rho_{eg}] - \Gamma \rho_{ee}\\
&\dot{\rho}_{gg} = \mathrm{Im}[\mu E(t) \rho_{eg}] + \Gamma \rho_{ee}\\
&\dot{\rho}_{eg} = i\frac{\mu E(t)}{2} (\rho_{ee} - \rho_{gg}) + \left(i\Delta - \frac{\Gamma}{2}\right) \rho_{eg}
\end{align*}
$\rho_{ee}$ and $\rho_{gg}$ represent the excited and ground state populations, $\rho_{eg}$ is the coherence term, $\mu$ is the dipole matrix element, $\Gamma$ is the spontaneous emission rate, and $\Delta = \omega_0 - \omega$ is the detuning between the comb carrier frequency and the transition frequency.

\subsubsection{Frequency comb source}

The EUV generation system can use a neodymium-doped yttrium aluminium garnet (Nd:YAG) laser \cite{cingoz_2012}, a titanium-sapphire ($Ti:Al_2O_3$) laser \cite{seres_2019}, or a mode-locked Yb-doped fiber laser \cite{zhang_2022}, \cite{pupeza_cavity_2014}.
% producing femtosecond pulses at a central wavelength of approximately 1030 nm.
Ytterbium-doped fiber lasers (Yb-fiber lasers) are widely used in modern extreme ultraviolet (EUV) generation systems, particularly in frequency comb spectroscopy applications. These lasers operate around 1030 to 1060nm and typically provide femtosecond pulses with durations of about 100fs at repetition rates in the range of 75 to 100~MHz. Their high stability and compatibility with fiber amplifier architectures make them ideal candidates for seeding femtosecond enhancement cavities used in high-harmonic generation (HHG). Due to their narrow linewidths and high average power, they are particularly suited for generating stable, tunable frequency combs in the vacuum ultraviolet (VUV) and EUV ranges. These systems are advantageous for high-resolution spectroscopy, such as probing the narrow nuclear transition of $^{229\text{m}}$Th.

Titanium-sapphire (Ti:sapphire) lasers are another important class of lasers used in EUV generation. Operating around 800nm, these lasers are capable of producing sub-30fs pulses with extremely high peak intensities. Although their typical repetition rates are lower than those of Yb-fiber systems (often in the kilohertz range for amplified systems or tens of megahertz for oscillators), they excel in generating isolated attosecond pulses through HHG in gas targets. Their broad spectral bandwidth and ultrashort pulse duration make them well-suited for time-resolved measurements at attosecond resolution. However, the large temporal bandwidth limits their utility in high-resolution spectroscopy due to the Fourier-transform limit.

Chromium-based lasers such as Cr:forsterite and Cr:YAG operate in the longer wavelength regime, typically from 1230 to 1300nm. These lasers are less commonly used in EUV applications but can be beneficial in extending the cutoff energy of high-order harmonics due to the favorable $\lambda^2$ scaling of the HHG process. However, their use is limited by lower conversion efficiencies and less mature laser technology compared to Yb-fiber and Ti:sapphire systems.

Nd:YAG and Nd:vanadate lasers, operating near 1064~nm, have been historically used in nonlinear optics schemes, such as resonant-enhanced four-wave mixing, to generate ultraviolet or VUV radiation. These lasers typically emit pulses in the picosecond to nanosecond regime and operate at lower repetition rates, making them less suitable for frequency comb generation or attosecond science. Their utility lies primarily in initial broadband spectroscopic surveys rather than high-precision metrology.

\begin{table}[h]
\centering
\begin{tabular}{ccccc}
\toprule
laser & wavelength & \begin{tabular}{@{}c@{}}pulse\\duration\end{tabular} & \begin{tabular}{@{}c@{}}repetition\\rate\end{tabular}\\ \midrule
Yb-Fiber            &1030nm       & $\sim$100 fs            & 75-100 MHz\\
Ti:Sapphire         &800nm        & $<$30 fs                & kHz-MHz\\
Cr:Forsterite       &1250nm       & fs                      & MHz\\
Nd:YAG              &1064nm       & ps-ns                  & kHz\\ \bottomrule
\end{tabular}
\caption{Comparison of laser systems used in EUV/VUV applications}
\end{table}

Each laser platform provides trade-offs between spectral resolution, temporal resolution, output power, and harmonic generation efficiency. Yb-fiber systems are optimal for precision measurements and nuclear clock development, while Ti:sapphire lasers remain the preferred option for attosecond time-domain studies.

\section{Optical cavities, quantum optics, and photonic engineering in solar cells}
\subsection{Optical cavities and light management in solar cells}
Optical cavities are resonant structures that confine light between reflective surfaces, allowing for enhanced light-matter interaction through multiple internal reflections. In solar cells, they can be used to enhance absorption in thin films. Embedding a thin photovoltaic (PV) layer within an optical cavity, such as a Fabry Perot or hemi-ellipsoidal cavity, allows incident light to bounce multiple times through the absorber, increasing the effective optical path length without increasing the physical thickness \cite{polman_2012}. This boosts absorption, especially in materials with low intrinsic absorption coefficients, such as indirect-gap semiconductors like silicon \cite{yablonovitch_1982}.

Optical cavities also offer a route to surpass the Yablonovitch limit, which is the theoretical absorption enhancement limit in isotropic media \cite{weinstein_2014}. By tailoring angular and spectral optical modes through nanophotonic design or metasurfaces, these cavities can promote favorable guided or angle-restricted modes that increase photon confinement \cite{weinstein_absorption_2015}. Furthermore, optical cavities help reduce radiative recombination losses by reflecting emitted photons back into the absorber \cite{betancur_2010}. This allows for photon recycling and raises the quasi-Fermi level splitting, thereby increasing the open-circuit voltage and overall efficiency.

% \subsection{Optical transistors}
Optical transistors can be implemented by systems based on electromagnetically induced transparency, microcavities with nonlinear media, and Rydberg atom-based setups. These devices are fundamental for developing all-optical logic gates and quantum information processing units. In the context of solar cells, optical transistors could be used to dynamically manage and redirect incoming sunlight, enhancing light trapping, spectrum splitting, or routing photons to specific subcells in multi-junction architectures for maximized absorption efficiency \cite{zehender_solar_trans_2021}.
Quantum optical switches, closely related to optical transistors, enable control over photon transmission using quantum states of matter. For instance, researchers have demonstrated a switch where a single atom can control the passage of a photon through a cavity, and vice versa. This interaction is pivotal for developing quantum networks and photonic information systems. When applied to solar technology, such quantum control elements could one day allow solar cells to respond adaptively to changing light conditions or even filter and convert photons with high precision, enabling next-generation intelligent or quantum-enhanced photovoltaic systems.

\subsection{Lasers and light concentration}

Lasers can be used as optical components in solar cells. Laser physics provides valuable tools and concepts for solar technology. In concentrator PV systems, coherent or quasi-coherent light sources like lasers or LEDs are used in device testing and calibration. Beam shaping and focusing techniques developed in laser optics enable precise delivery of concentrated light onto high-efficiency PV cells, such as multi-junction GaAs. Stimulated emission concepts from laser physics also inform high-performance solar cells. In materials such as GaAs and perovskites, radiatively emitted photons can be reabsorbed within the device, enhancing external luminescence and boosting the open-circuit voltage, analogous to stimulated emission in lasers. Additionally, lasers are integral to PV manufacturing, where they are used for scribing, texturing, doping, and annealing. These processes benefit from the spatial precision and minimal thermal damage offered by laser-based techniques.

\begin{figure}
\includegraphics[width = 0.45\textwidth, height = 0.3\textwidth]{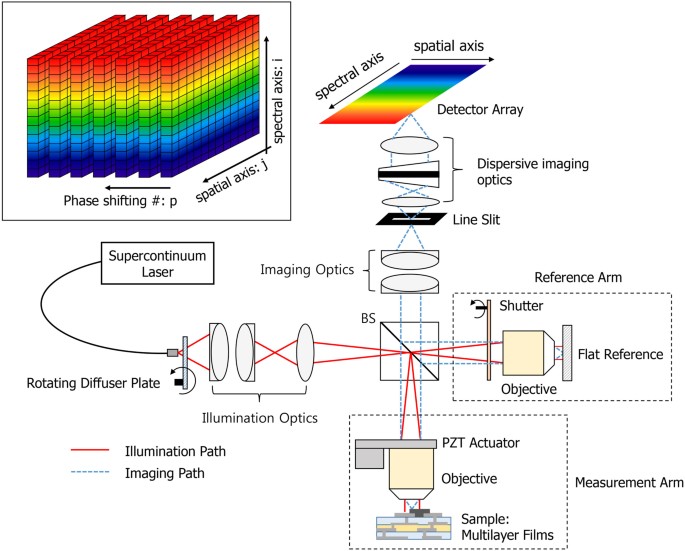}
\caption{schematic diagram of a Linnik-type interferometric setup using spectrally-resolved white-light interferometry for 3D thickness measurement of multilayer thin-film structures \cite{ghim_simultaneous_film_2017}}
\label{simultaneous film surfaces, ghim, 2017}
\end{figure}

Figure \ref{simultaneous film surfaces, ghim, 2017} illustrates a schematic diagram of the proposed measurement technique for obtaining the three-dimensional thickness profile of a multilayer film structure \cite{ghim_simultaneous_film_2017}.
This technique is motivated by the widespread use of transparent conductive thin-film layers in multilayer circuits, which are critical components in industries such as semiconductors, flat panel displays, and light-emitting diodes (LEDs).
% The development of three-dimensional packaging technologies in the semiconductor sector has driven the need for volumetric inspection methods capable of resolving internal structural features.
Organic light-emitting diode (OLED) displays, which employ multilayer transparent electrodes, and photovoltaic devices, require precise metrology of internal thin films to optimize manufacturing processes and product performance.
The measurement system is based on a Linnik interferometric configuration combined with spectrally-resolved white-light interferometry. A broadband supercontinuum laser source is employed to generate a wide spectral range of illumination.
% Although such a high-coherence source typically introduces speckle noise due to its spatial coherence, this is effectively suppressed using a rotating diffuser plate, allowing for uniform illumination of the sample.
Kohler illumination is utilized to produce a uniformly lit field of view across two identical microscope objectives arranged in a Linnik-type interferometric setup.

The system comprises of the reference arm and the measurement arm. In the reference arm, a flat mirror generates the reference wave, which can be selectively blocked or transmitted using a controllable optical shutter.
% The measurement wave is produced by reflecting light from the sample under investigation in the measurement arm.
% Depending on the state of the shutter, the system captures either the sample's reflected beam alone or the interference between the sample and the reference wave.
% The optical paths in both arms are imaged onto an entry line slit through imaging optics, which maps the field of view of each microscope objective. The line slit, in conjunction with dispersive imaging optics, enables a two-dimensional detector array to capture spatially resolved spectral information. One axis of the detector corresponds to spatial position along a selected line on the sample, while the other axis records the spectral content.
To extract precise phase information across the broadband spectrum, a piezoelectric actuator (PZT) in the measurement arm introduces controlled phase shifts between successive measurements. Instead of requiring exact phase shifts matched to the wavelength of the light source, which is challenging for broadband illumination, the system employs an iterative least-squares phase-shifting algorithm. This algorithm assumes arbitrary phase shifts and retrieves accurate phase information from several phase-shifted spectral interferograms, enabling robust reconstruction of the multilayer film structure over a wide spectral range.

In the 2D detector array in figure \ref{simultaneous film surfaces, ghim, 2017}, the measured intensity $I_{ijp}$ at pixel $(i,j)$ with phase shift $\delta_p$ is
\begin{equation*}
I_{ijp} = D_{ij} + V_{ij} \cos(\Phi_{ij} - \delta_p)
\end{equation*}
where $D_{ij} = I^{\text{ref}}_{ij} + I^{\text{mea}}_{ij}$ and $V_{ij} = 2 \sqrt{I^{\text{ref}}_{ij} I^{\text{mea}}_{ij}}$.
Assuming $\delta_1 = 0$, define the intensity difference:
\begin{equation*}
\eta_{ijp} = I_{ijp} - I_{ij1} = C_{ij} \cos \delta_p + S_{ij} \sin \delta_p
\end{equation*}
where $C_{ij} = V_{ij} \cos \Phi_{ij}$ and $S_{ij} = V_{ij} \sin \Phi_{ij}$.
Minimizing the least-squares error function:
\begin{equation*}
E_{ij} = \sum_{p=1}^m (\eta_{ijp} - C_{ij} \cos \delta_p - S_{ij} \sin \delta_p)^2
\end{equation*}
yields the solution for $C_{ij}$ and $S_{ij}$ via matrix inversion, leading to the phase $\Phi_{ij} = \tan^{-1} \left( \frac{S_{ij}}{C_{ij}} \right)$.
To iteratively refine unknown $\delta_p$, define and solve
\begin{align*}
&E_p = \sum_{j=1}^l (\eta_{ijp} - C_{ij} \cos \delta_p - S_{ij} \sin \delta_p)^2\\
&\delta_p = \tan^{-1} \left( \frac{\sum_j \eta_{ijp} \sin \Phi_{ij}}{\sum_j \eta_{ijp} \cos \Phi_{ij}} \right)
\end{align*}
The total phase for multilayer films is:
\begin{equation*}
\Phi(h, d_1 \dots d_n; k) = 2 k h + \psi(d_1 \dots d_n; k)
\end{equation*}
where $\psi$ includes a linear term $\psi_l$ and nonlinear term $\psi_{nl}$. The nonlinear part is isolated as $\Phi_{nl}(d_1 \dots d_n; k) = \Phi - 2kh$.
Spectral reflectance is measured using relative reflectance
\begin{equation*}
\mathcal{R}_{\text{sam}}(k) = \frac{I_{\text{sam}}(k)}{I_{\text{std}}(k)} \mathcal{R}_{\text{ref}}(k)
\end{equation*}
with $\mathcal{R}_{\text{ref}}(k)$ computed from Fresnel equations:
\begin{equation*}
\mathcal{R}_{\text{ref}}(k) = \left( \frac{n(k)-1)^2 + \kappa(k)^2}{(n(k)+1)^2 + \kappa(k)^2} \right)
\end{equation*}
A merit function combines phase and reflectance to retrieve film thicknesses
\begin{align*}
&\xi(d_1 \dots d_n)=\sum_i \alpha [\Phi_{nl}^E(k_i) - \psi_{nl}^T(d_1 \dots d_n; k_i)]^2\\
&\quad+ \chi [\mathcal{R}_E(k_i) - \mathcal{R}_T(d_1 \dots d_n; k_i)]^2
\end{align*}
Using known refractive indices, the thicknesses $d_1 \dots d_n$ are optimized by minimizing $\xi$.
with known $d_i$, surface height $h$ is retrieved by minimizing:
\begin{equation*}
\Gamma(h) = \sum_i [\Phi^E(k_i) - \psi^T(d_1 \dots d_n; k_i) - 2 k_i h]^2
\end{equation*}

\subsection{Nanophotonics and optical components for solar enhancement}

Quantum optics, which treats light as quantized photons and explores phenomena such as coherence and entanglement, contributes to emerging concepts in photovoltaics. One avenue is photon upconversion and downconversion, where quantum-engineered materials convert two low-energy photons into one high-energy photon or vice versa. These mechanisms aim to better utilize photons that would otherwise be below or above the absorber’s bandgap, increasing theoretical efficiency limits. Other experimental efforts explore the use of entangled photon pairs and coherent exciton transport to reduce thermalization and recombination losses. Inspired by quantum coherence observed in photosynthetic systems, researchers investigate whether coherent superpositions can allow excitons to travel with minimal losses, offering potential advantages in organic or hybrid solar materials.

Incorporating advanced optical components and nanophotonics into solar cell architectures enables superior light trapping and management. Photonic crystals and metasurfaces with engineered periodicity can manipulate the propagation of light, enhance interaction times, or concentrate specific wavelengths into the absorber layer. These structures can also support slow-light modes or guided resonances that increase absorption without requiring thicker active layers. Plasmonic nanostructures composed of metals like silver or gold can concentrate electromagnetic fields at the nanoscale, boosting local absorption in adjacent semiconductors. Interference coatings and Bragg reflectors are widely used to minimize reflection losses or reflect unabsorbed light back into the device. Angular and spectral filters further tailor the incoming solar spectrum and suppress unwanted emission, contributing to both thermal and electrical optimization.

% \subsection{Hybrid and cavity-enhanced architectures}

Cavity-enhanced architectures offer promising directions for improving thin-film solar cell performance. For example, organic solar cells that integrate metallic electrodes can form Fabry--Pérot cavities that enhance photoluminescence and suppress photon escape. This results in improved charge collection and voltage. Perovskite solar cells, with their high photoluminescence efficiency, are well suited for optical resonators such as distributed Bragg reflectors, which can enhance light emission and recycling. In luminescent solar concentrators, fluorescent dyes embedded in optical waveguides absorb incident light and re-emit it isotropically. The emitted light undergoes total internal reflection and is guided to the edges of the device, where it is absorbed by PV cells. These concentrators benefit from optical design that enhances trapping and directionality, increasing efficiency without requiring solar tracking systems.

\section{Material Performance Analysis}
The need for high‐efficiency, low‑cost solar energy conversion demands novel materials and architectures. Organic/inorganic hybrid semiconductors, combining the tunable bandgaps of molecular systems with the robustness of inorganic crystals, offer a pathway to tandem devices that can exceed conventional efficiency limits.
% It is crucial to establish and refine fabrication and synthesis methods for optoelectronic materials, and to apply these methodologies toward the development of next‑generation photovoltaic devices and ultra‑sensitive photodetection platforms.
The need for renewable energy sources catalyzed the development of advanced photovoltaic technologies. Semiconductor solar cells, perovskites \cite{li_multidim_2021}, organic materials \cite{liu_organic_2022}, \cite{zuccala_2023}, transition metal dichalcogenides (TMDs) \cite{nazif_2021}, cadmium telluride (CdTe), and silicon have emerged as leading candidates due to their distinct material properties and potential for high efficiency. Perovskite solar cells (PSCs) have rapidly achieved high power conversion efficiencies (PCEs) with inverted p-i-n architectures showing improved stability \cite{liu_progress_2023}. Despite this progress, long-term operational stability and lead toxicity remain major barriers to commercialization. TMD-based solar cells, such as those employing WSe$_2$, offer advantages like ultrathin flexibility and high specific power, making them suitable for aerospace applications \cite{nguyen_2023}. However, they are hindered by low efficiency and complex fabrication methods. CdTe solar cells are among the most commercially successful thin-film technologies, benefiting from a direct bandgap and high absorption coefficients. Yet, environmental concerns due to cadmium toxicity and resource limitations challenge their long-term sustainability. Silicon remains the dominant photovoltaic material, with mature manufacturing infrastructure and reliable long-term performance, but it is approaching its theoretical efficiency limit, spurring interest in tandem configurations with perovskites.

Experimental characterization techniques, such as angle-resolved photoemission spectroscopy (ARPES), X-ray diffraction (XRD), and scanning tunneling microscopy (STM) can be used to characterize the crystallographic and electronic structures of photovoltaic materials. ARPES provides direct access to the band structure of thin-film semiconductors, offering valuable data on band alignment and dispersion. XRD reveals the phase purity and crystallinity of perovskite and chalcogenide materials, which are directly correlated with device performance and stability. These tools are indispensable for optimizing the microstructure and phase behavior of active layers in solar cells. Condensed matter physics provides the theoretical foundation necessary to understand and manipulate the electronic, optical, and structural properties of materials used in photovoltaic devices. Key theories such as band structure, carrier transport, exciton dynamics, and defect physics underpin the operation of solar cells. Techniques derived from condensed matter theory, including effective mass approximations, k·p perturbation theory, and many-body methods like Hartree Fock theory and Bethe-Salpeter Equation (BSE), are used to predict and control the electronic and optical responses of semiconductor materials. These insights are crucial for designing absorber layers with optimal bandgaps, minimal recombination losses, and high charge carrier mobilities.

Nanofabrication methods in materials science are essential to complement these approaches by focusing on synthesis, processing, and property-performance relationships. Thin-film deposition methods such as sputtering, chemical vapor deposition (CVD), atomic layer deposition (ALD), and solution-processing techniques like spin-coating and blade coating enable precise control over film thickness, composition, and interface quality \cite{friedman_2017}. Interface engineering plays a crucial role in minimizing charge recombination and achieving favorable energy-level alignments in multi-layered device architectures \cite{kim_2024}. These methods allow researchers to fabricate heterostructures with tailored band offsets and optimized interfacial properties for efficient carrier extraction \cite{aftab_2023}. Defect engineering addresses non-radiative recombination centers and long-term device degradation \cite{kozlov_2025}. By identifying and passivating deep-level traps, researchers can extend the carrier lifetimes and operational lifetimes of devices. Techniques such as deep-level transient spectroscopy (DLTS), photoluminescence spectroscopy, and Kelvin probe force microscopy (KPFM) help in understanding defect types, densities, and distributions. These characterizations inform the synthesis of high-purity materials and the design of effective passivation strategies. The mechanical, thermal, and environmental stability of solar cell materials is deeply rooted in materials science. The study of grain boundaries, thermal expansion coefficients, and mechanical flexibility informs the selection and integration of materials for applications requiring durability and lightweight form factors, such as wearable or space-based photovoltaics.

\subsection{Perovskite solar cells}
\subsubsection{Photovoltaic process}
Perovskite solar cells (PSC), particularly those based on methylammonium lead iodide (MAPbI$_3$), have achieved power conversion efficiencies (PCE) exceeding 26\% \cite{liu_progress_2023}. The crystalline structure of perovskites significantly influences their optical and electronic properties. For instance, the cubic phase of MAPbI$_3$ exhibits a higher bandgap ($\sim$1.7 eV) and enhanced optical absorption, leading to improved device performance \cite{li_multidim_2021}. Studies have shown that PSCs with cubic MAPbI$_3$ can achieve efficiencies up to 26\% with absorber layer thicknesses around 900 nm.

Integration of resonant photonic structures, such as photonic crystals and metasurfaces, into PSCs has been explored to enhance light absorption and photocurrent generation \cite{aftab_2023}. By incorporating these structures, researchers have achieved an efficiency of 24.4\% in resonant solar cells, demonstrating the potential of combining perovskite materials with resonant photonics.

\begin{figure}
\includegraphics[width = 0.45\textwidth, height = 0.2\textwidth]{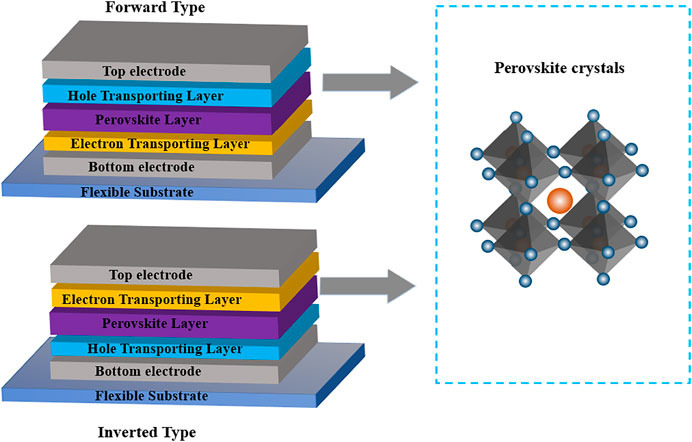}
\caption{perovskite device structure \cite{zheng_chen_process_2021}}
\label{zheng_chen_process_2021}
\end{figure}

In a perovskite solar cell, the photovoltaic process begins with photon absorption by the perovskite absorber layer (see figure \ref{zheng_chen_process_2021}), commonly composed of hybrid organic-inorganic lead halide compounds such as methylammonium lead iodide (CH$_3$NH$_3$PbI$_3$) \cite{long_hu_quantum_dot_2021}. These materials typically have direct optical bandgaps ranging from approximately 1.55 eV to 2.3 eV, depending on the halide composition (X = I, Br, or Cl), making them highly efficient at absorbing visible light \cite{ying_shang_2025}. When a photon with energy equal to or greater than the bandgap is absorbed, an electron is excited from the valence band to the conduction band, leaving behind a hole. In CH$_3$NH$_3$PbI$_3$, the valence band is primarily derived from iodine 5$p$ orbitals, while the conduction band is dominated by lead 6$p$ orbitals. This interband transition results in the creation of an electron-hole pair.

\begin{figure}
\includegraphics[width = 0.45\textwidth, height = 0.2\textwidth]{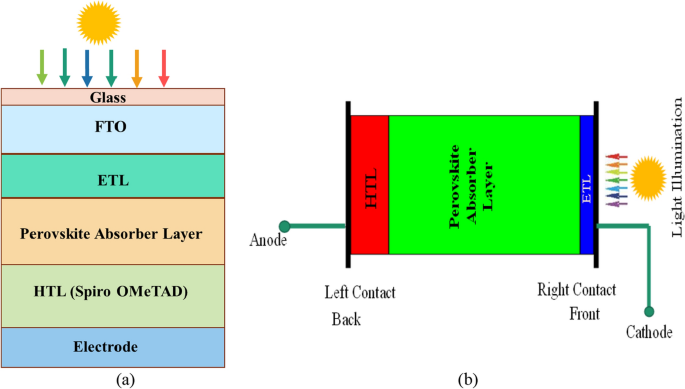}
\caption{perovskite absorber layer, electron and hole transport layers \cite{kumar_heterojunction_2023}}
\label{kumar heterojunction 2023}
\end{figure}

Unlike many organic photovoltaic materials, perovskites exhibit very low exciton binding energies (typically less than 25 meV), which is comparable to or less than the thermal energy at room temperature. As a result, photogenerated excitons in perovskites can spontaneously dissociate into free charge carriers without the need for a donor-acceptor heterojunction. This intrinsic property greatly simplifies the device architecture and enhances charge separation efficiency. Once separated, the electrons and holes are selectively extracted by adjacent transport layers. The electrons migrate through the electron transport layer (ETL) (see figure \ref{kumar heterojunction 2023}), which is often composed of materials such as titanium dioxide (TiO$_2$) or tin dioxide (SnO$_2$). These materials possess conduction bands well-aligned with that of the perovskite, enabling efficient electron extraction and transport toward the cathode.

Simultaneously, the holes are extracted through the hole transport layer (HTL), which may consist of organic small molecules such as Spiro-OMeTAD or polymeric alternatives like PTAA. The HTL is engineered to have an appropriate highest occupied molecular orbital (HOMO) level aligned with the valence band of the perovskite to facilitate hole transfer while blocking electron flow, thereby minimizing recombination losses. The combined action of these layers ensures that charge carriers are efficiently collected at the respective electrodes, electrons at the cathode and holes at the anode, thereby generating an electric current. Device performance is strongly influenced by the quality of the perovskite film, interfacial energy-level alignment, and defect passivation strategies, all of which affect charge carrier lifetimes, mobilities, and recombination dynamics. The exceptional optoelectronic properties of perovskite materials, such as high absorption coefficients, long diffusion lengths, and tunable bandgaps, contribute to their rapidly rising prominence in next-generation photovoltaic technologies.

\subsubsection{Material characterization, screening, and fabrication process}
\begin{figure}
\includegraphics[width = 0.45\textwidth, height = 0.3\textwidth]{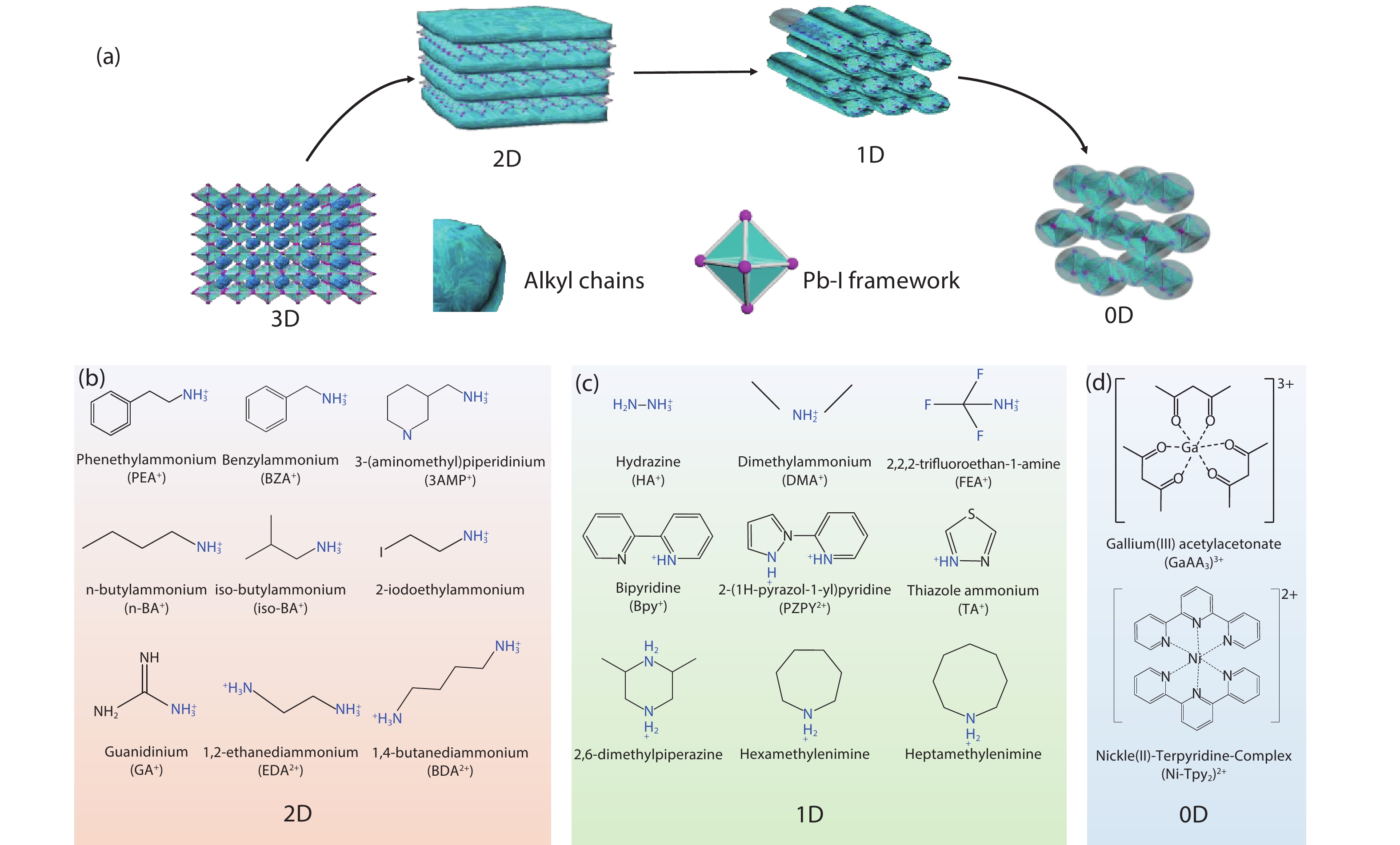}
\caption{lead iodide perovskite semiconductors, \cite{li_multidim_2021}}
\label{multidimensional perovskites, li}
\end{figure}

Figure \ref{multidimensional perovskites, li} illustrates the structural features and chemical components of hybrid lead iodide perovskites with varying dimensionalities \cite{li_multidim_2021}. In panel (a), the crystal structures of 0D, 1D, 2D, and 3D perovskites are shown based on the connectivity of the $[MX_6]^{4-}$ octahedra. In 0D perovskites, the octahedra are completely isolated by surrounding organic cations. In 1D structures, the octahedra are connected in linear chains through face-sharing, edge-sharing, or corner-sharing configurations. In 2D perovskites, the octahedra form layered or corrugated sheets separated by large organic cations, whereas in 3D perovskites, the octahedra form a continuous three-dimensional network. It is important to note that this structural classification is distinct from the morphological definitions such as 0D nanoparticles, 1D nanowires, and 2D nanosheets. Panels (b-d) depict the chemical structures of A-site organic cations commonly used in low-dimensional (LD) and 3D perovskite solar cells. These organic spacers play a critical role in determining the dimensionality, phase stability, and moisture resistance of the resulting perovskite films.
% By varying the type and size of the organic ammonium cation, researchers can tune the structure and optoelectronic properties of the material to suit specific device applications.

\begin{figure}
\includegraphics[width = 0.45\textwidth, height = 0.3\textwidth]{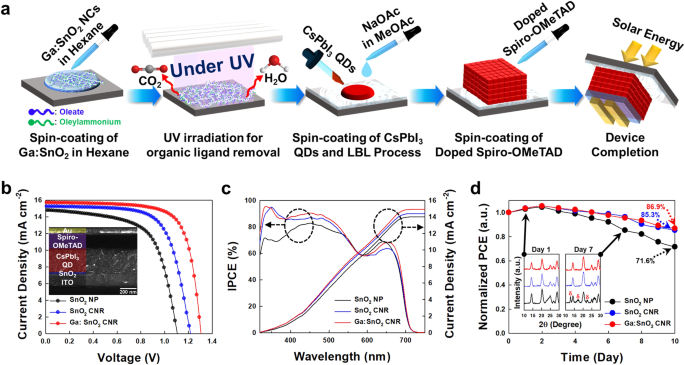}
\caption{fabrication process for perovskite quantum dot (PQD) light-emitting diodes (LEDs), encompassing the spin-coating of Ga-doped SnO$_2$, UV irradiation for organic ligand removal, layer-by-layer deposition of CsPbI$_2$ quantum dots, spin-coating of doped spiro-OMeTAD, and device completion \cite{kim_2024}}
\label{perovskite quantum dot, kim, 2024}
\end{figure}

As shown in figure \ref{perovskite quantum dot, kim, 2024}, a colloidal solution of gallium-doped tin dioxide (Ga:SnO\textsubscript{2}) nanocrystals is prepared in hexane \cite{kim_2024}. This solution is spin-coated onto a pre-cleaned indium tin oxide (ITO) substrate at 3000 rpm for 60 seconds, forming a uniform electron transport layer (ETL). The Ga doping adjusts the energy levels of SnO\textsubscript{2}, facilitating better alignment with the perovskite quantum dots and enhancing electron extraction efficiency.
% \subsection*{2. UV Irradiation for Organic Ligand Removal}
Post-deposition, the Ga:SnO\textsubscript{2} film undergoes ultraviolet (UV) irradiation to remove organic ligands from the nanocrystal surfaces. This treatment improves the film's conductivity and ensures better charge transport by eliminating insulating organic residues.
% \subsection*{3. Layer-by-Layer Deposition of CsPbI\textsubscript{2} Quantum Dots}
CsPbI\textsubscript{2} quantum dots (QDs) are synthesized and dispersed in a nonpolar solvent such as hexane. The QD solution is spin-coated onto the Ga:SnO\textsubscript{2} layer at 2000 rpm for 30 seconds. After each QD layer deposition, the film is treated with methyl acetate (MeOAc) to remove native oleate ligands, rendering the layer insoluble to subsequent coatings. This layer-by-layer (LbL) process is repeated multiple times to achieve the desired film thickness and uniformity, ensuring efficient charge transport and emission properties.
% \subsection*{4. Spin-Coating of Doped Spiro-OMeTAD Hole Transport Layer}
A hole transport layer (HTL) solution is prepared by dissolving spiro-OMeTAD in chlorobenzene, with additives such as lithium bis(trifluoromethanesulfonyl)imide (Li-TFSI) and 4-tert-butylpyridine (tBP) to enhance conductivity and stability. This solution is spin-coated onto the CsPbI\textsubscript{2} QD film at 4000 rpm for 30 seconds, forming a uniform HTL that facilitates hole extraction and transport.
% \subsection*{5. Device Completion}
A metal electrode, typically gold (Au), is thermally evaporated onto the spiro-OMeTAD layer under high vacuum conditions to complete the perovskite QD LED with layers: ITO/Ga:SnO\textsubscript{2}/CsPbI\textsubscript{2} QDs/Spiro-OMeTAD/Au.

\begin{figure}
\includegraphics[width = 0.45\textwidth, height = 0.3\textwidth]{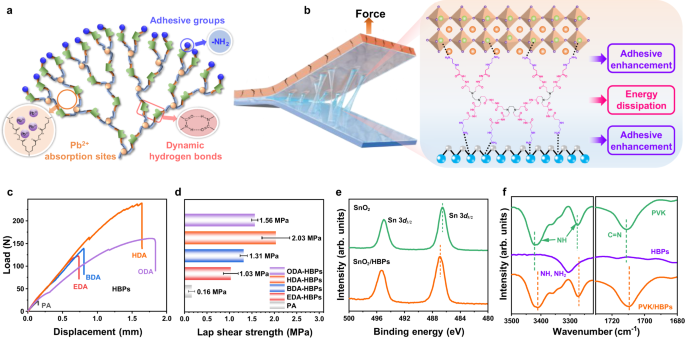}
\caption{structural design, interfacial bonding mechanism, and adhesive performance of hyperbranched polymers (HBPs) at the SnO$_2$/perovskite interface for enhanced mechanical stability and efficiency in perovskite solar cells \cite{zhihao_li_hyperbranch_2023}}
\label{hyperbranched perovskite, zhihao li, 2023}
\end{figure}

Figure \ref{hyperbranched perovskite, zhihao li, 2023} illustrates the molecular design, interfacial interactions, and functional benefits of incorporating hyperbranched polymers (HBPs) as adhesive interlayers between SnO$_2$ electron transport layers (ETLs) and perovskite active layers in both rigid and flexible perovskite solar cells.
Figure \ref{hyperbranched perovskite, zhihao li, 2023}a presents the molecular structure of the synthesized HBPs, which are based on polyamide-amine networks. These polymers are generated via a one-pot Michael addition reaction between N,N'-methylene diacrylamide (MBA) and various diamines. The selected diamines, ethylenediamine (EDA), 1,4-butanediamine (BDA), 1,6-hexanediamine (HDA), and 1,8-octanediamine (ODA), contribute different alkyl spacer lengths, thereby tuning the flexibility and mechanical properties of the resulting HBPs. The polymers exhibit a hyperbranched, globular topology featuring dense surface functionalities, including primary amine (-NH$_2$), secondary amine (-NH-), carboxyl (-COOH), and amide (C=O) groups, which play a critical role in bonding and mechanical damping.

\begin{figure}
\includegraphics[width = 0.3\textwidth, height = 0.2\textwidth]{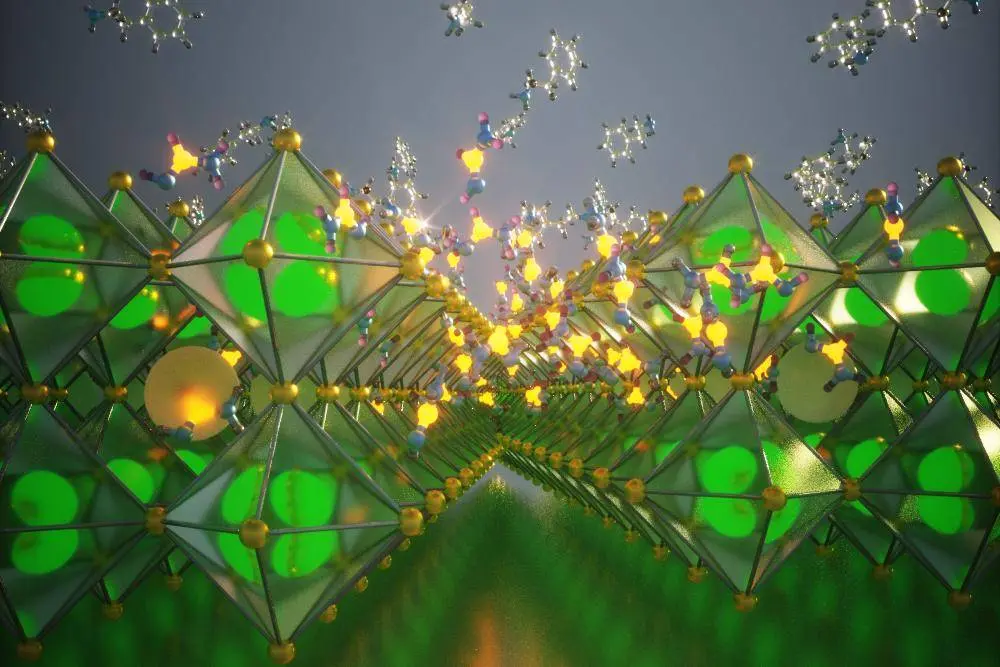}
\caption{experiment that analyzes the degradation mechanisms of mixed-cation metal halide perovskites, particularly Cs$_x$FA$_{1-x}$PbI$_3$, under humid conditions \cite{hidalgo_synergistic_2023}}
\label{hidalgo_synergistic_2023}
\end{figure}

Mixed-cation metal halide perovskites are known for their high power conversion efficiencies exceeding 26\%, are limited by their structural instability in the presence of water and oxygen, shown in figure \ref{hidalgo_synergistic_2023}. Using in situ grazing incidence wide-angle X-ray scattering (GIWAXS), the authors monitored phase evolution during exposure to H$_2$O/air, H$_2$O/N$_2$, and dry air \cite{hidalgo_synergistic_2023}. The results revealed that exposure to H$_2$O/air led to the transformation of the tetragonal perovskite $\beta$-phase (space group $P4/mbm$) into two nonperovskite phases: hexagonal FAPbI$_3$ (2H, space group $P6_3/mmc$) and orthorhombic CsPbI$_3$ ($\delta$-Cs, space group $Pnma$). These transformations were significantly slower or absent when either H$_2$O or O$_2$ was removed, highlighting a synergistic degradation pathway involving the dissolution of FAI by water and subsequent oxidation of iodide by oxygen.

\subsubsection{Computational material simulations}
\begin{figure}
\includegraphics[width = 0.45\textwidth, height = 0.25\textwidth]{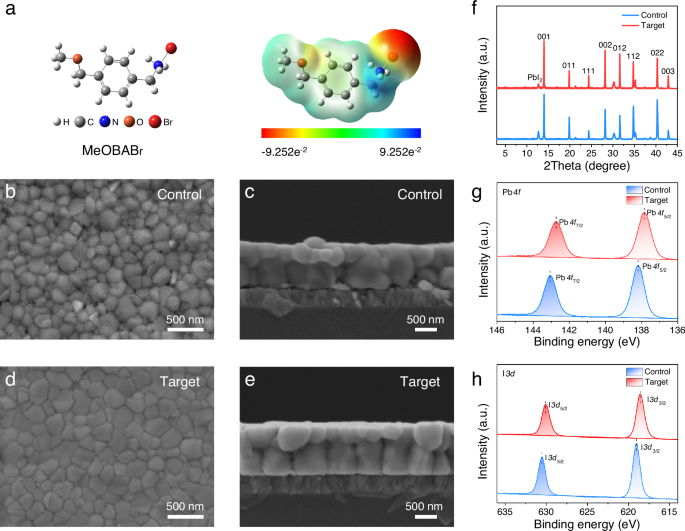}
\caption{MeOBABr serves multiple functional roles: it modulates the nucleation and growth of perovskite crystals by interacting with \(\text{Pb}^{2+}\), enhances film morphology by increasing grain size and uniformity, and passivates electronic defects. (a). electrostatic potential profile (ESP), (b-e). scanning electron microscopy (SEM), (f). X-ray powder diffraction (XRD), (g,h). X-ray photoelectron spectroscopy (XPS)
 \cite{jun_jin_spontaneous_bifacial_2025}}
\label{spontaneous bifacial capping perovskite, jin}
\end{figure}

Figure \ref{spontaneous bifacial capping perovskite, jin} provides a comprehensive set of characterizations that illustrate how the organic ammonium salt 4-(methoxy)benzylamine hydrobromide (MeOBABr) influences the microstructure, crystallinity, and electronic structure of perovskite films \cite{jun_jin_spontaneous_bifacial_2025}. The motivation for incorporating MeOBABr stems from its unique molecular structure, particularly its ability to interact strongly with \(\text{Pb}^{2+}\) ions through ionic bonding, which alters the crystallization dynamics of the perovskite material.
Panel (a) shows the molecular structure of MeOBABr alongside its calculated electrostatic potential (ESP) map. The negative charge density is concentrated near the bromide ion, a feature attributed to the electron-donating methoxy group on the aromatic ring. This charge distribution suggests that MeOBABr can strongly interact with positively charged lead ions in the perovskite precursor solution. This interaction is quantitatively supported by binding energy calculations, which reveal that MeOBABr binds more strongly to \(\text{Pb}^{2+}\) than conventional solvents like dimethylformamide (DMF) or dimethyl sulfoxide (DMSO). This stronger binding alters the colloidal chemistry of the precursor solution, promoting the formation of larger colloidal clusters, as evidenced by dynamic light scattering (DLS) measurements. Larger colloidal clusters lead to reduced nucleation density and thereby more uniform and larger crystal domains in the final film. Panels (b) and (c) display top-view and cross-sectional scanning electron microscopy (SEM) images of the control perovskite film prepared without MeOBABr. These images reveal small grains with an average diameter of about 200 nm and a heterogeneous distribution in grain size. The film surface appears rough and less compact, which could negatively impact charge transport and film stability.

\begin{figure}
\includegraphics[width = 0.45\textwidth, height = 0.25\textwidth]{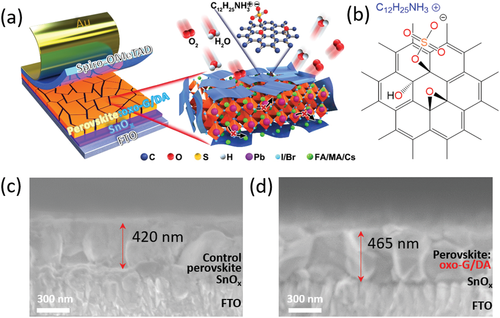}
\caption{(a). structural schematics, (b). chemical composition, (c,d). cross-sectional scanning electron microscopy (SEM) morphology of perovskite solar cells with and without oxo-G/DA nanosheets, showing improved crystallization and film quality upon oxo-G/DA incorporation \cite{meng_li_ultrathin_nanosheets_2019}}
\label{ultrathin nanosheets peroskite, meng li, 2019}
\end{figure}

Figure \ref{ultrathin nanosheets peroskite, meng li, 2019} presents the structural and morphological characterization of perovskite solar cells (PSCs) enhanced with ultrathin 2D nanosheets of oxo-functionalized graphene modified by dodecylamine (oxo-G/DA). This novel hybrid material is used to improve both the efficiency and stability of PSCs by influencing the crystallization process and interfacial properties of the perovskite layer.
Figure \ref{ultrathin nanosheets peroskite, meng li, 2019}a illustrates the schematic device architecture of the fabricated PSC, which adopts a planar n-i-p configuration. The layers from bottom to top are fluorine-doped tin oxide (FTO), tin oxide electron transport layer (SnO$_x$), a mixed halide perovskite absorber layer incorporating oxo-G/DA, hole transport layer (Spiro-OMeTAD), and a gold electrode. On the right of Figure 1a is a schematic simulation of the microstructure showing how ultrathin oxo-G/DA nanosheets are uniformly embedded within the perovskite crystal matrix. These nanosheets are well-dispersed due to the presence of dodecylamine (DA), which enhances their solubility and reduces aggregation. The incorporation of oxo-G/DA leads to the encapsulation of nanosheets within the growing perovskite grains during the crystallization process.

\subsection{Organic solar cells}
In organic solar cells, the photon absorption process begins when a photon is absorbed by the organic donor material, such as a polymer like poly(3-hexylthiophene-2,5-diyl) (P3HT). This absorption excites an electron from the highest occupied molecular orbital (HOMO), which is analogous to the valence band in semiconductors, to the lowest unoccupied molecular orbital (LUMO), which is comparable to the conduction band. In organic molecules, the pi electrons are primarily localized within individual molecules or polymer segments, leading to the formation of discrete molecular orbitals. The absorbed energy creates a bound exciton, which is an electron-hole pair. However, at this stage, the electron and hole remain bound together and are not yet separated into free charge carriers. 

The exciton generated in the donor material then begins to diffuse toward the interface with the acceptor material. This diffusion typically occurs over a few nanometers, with an average distance of less than 10 nm, ensuring the exciton reaches the donor-acceptor interface where charge separation takes place. The acceptor material, often a fullerene derivative like phenyl-C61-butyric acid methyl ester (PCBM), plays a critical role in facilitating the separation of the electron-hole pair. At the donor-acceptor interface, the electron is transferred from the LUMO of the donor to the LUMO of the acceptor, as the acceptor material has a lower LUMO energy level than the donor. The hole remains localized on the donor material, while the electron is now effectively "released" as a free carrier onto the acceptor material. This process marks the point where charge separation occurs, and the electron is now free to move within the device.

Once the electron is separated from the hole, it moves through the acceptor material's network toward the cathode, while the hole migrates through the donor material toward the anode. This separation and transport of charge carriers are essential for generating the photocurrent in organic solar cells. The efficiency of the solar cell depends on several factors, including the distance the exciton must diffuse to reach the interface, the energy alignment between the donor and acceptor materials, and the ability of the charge carriers to transport through the respective materials without recombination. Ultimately, the effective collection of these charge carriers at the electrodes, combined with minimal losses due to recombination, defines the power conversion efficiency of the organic solar cell.

\begin{figure}
\includegraphics[width = 0.45\textwidth, height = 0.27\textwidth]{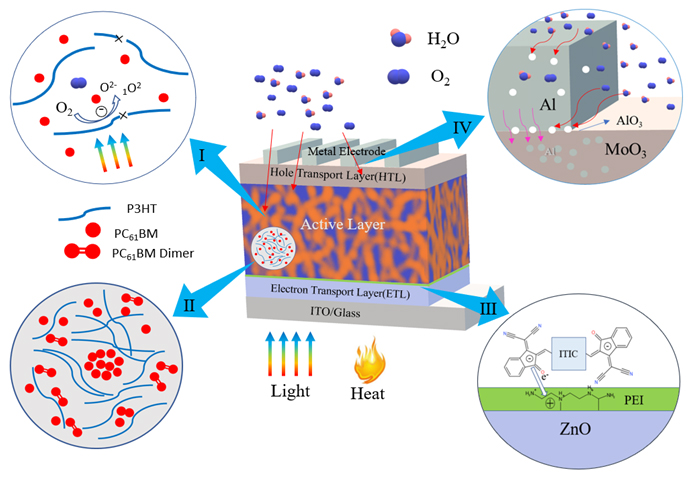}
\caption{a schematic diagram of an inverted organic solar cell device structure, along with four typical degradation pathways that can reduce device performance over time \cite{liu_organic_2022}}
\label{yanfu liu, degradation organic, 2022}
\end{figure}

Figure \ref{yanfu liu, degradation organic, 2022} presents a schematic diagram of an inverted organic solar cell (OSC) structure, highlighting four typical degradation pathways that lead to performance deterioration. In an inverted OSC configuration, the device consists of a substrate, typically glass or flexible plastic, coated with an indium tin oxide (ITO) electrode, followed by an electron transport layer (ETL) such as ZnO or TiO$_x$, an active layer composed of a donor and a non-fullerene acceptor (commonly from the Y-series), a hole transport layer (HTL) such as MoO$_3$ or PEDOT:PSS, and finally a top electrode, often silver (Ag) or aluminum (Al). The degradation pathways identified in the figure are: (I) molecular photo-oxidation, in which exposure to light and oxygen leads to chemical degradation of the active layer materials, particularly the oxidation of donor or acceptor molecules, thereby reducing charge transport efficiency; (II) aggregation and fullerene dimerization under photothermal conditions, where thermal stress induces morphological changes such as phase segregation or fullerene-based dimer formation that disrupt the nanoscale heterojunction necessary for charge separation; (III) interfacial reactions occurring at the boundaries between the active layer and the charge transport layers or electrodes, which can result in increased series resistance and trap formation; and (IV) water and oxygen erosion combined with electrode metal diffusion, where moisture and oxygen ingress can corrode metal electrodes or diffuse into the device, leading to irreversible chemical and physical damage.
% These degradation mechanisms collectively illustrate the multifactorial nature of OSC instability and underscore the importance of addressing both intrinsic material vulnerabilities and extrinsic environmental stressors.

\begin{figure}
\includegraphics[width = 0.45\textwidth, height = 0.27\textwidth]{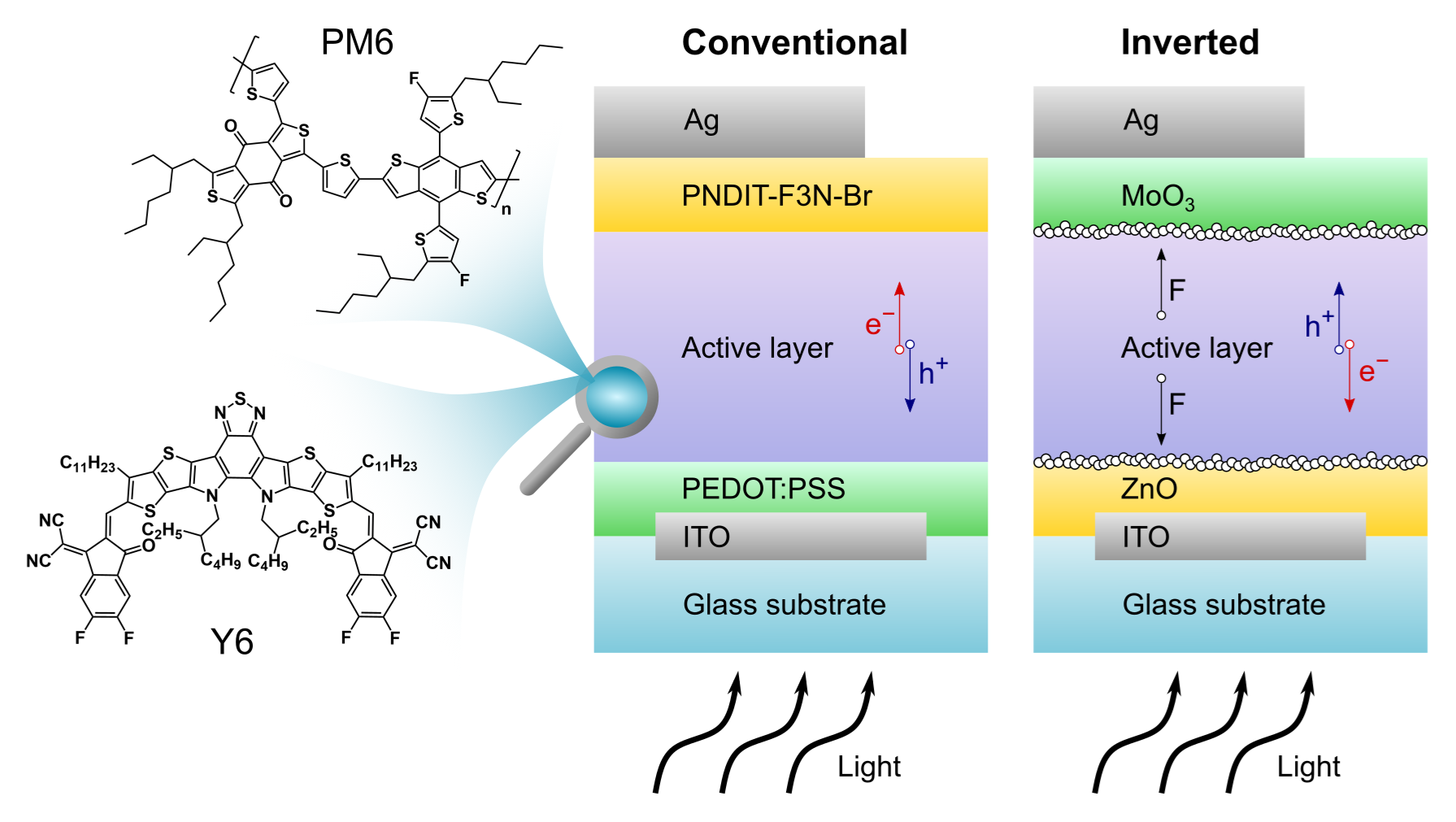}
\caption{PM6 (wide bandgap polymer donor) and Y6 (acceptor) molecules, layer structure of conventional and inverted organic solar cells \cite{zuccala_2023}}
\label{elena zuccala, halogens absorber, 2023}
\end{figure}

Figure \ref{elena zuccala, halogens absorber, 2023} illustrates three components: the molecular structures of PM6 and Y6, the layer structure of a conventional organic solar cell, and the layer structure of an inverted organic solar cell. PM6 is shown as a conjugated polymer containing fluorine atoms along its backbone to enhance molecular packing and energy level alignment, while Y6 is represented as a small-molecule non-fullerene acceptor with a fused-ring core and halogenated end groups, such as chlorine or fluorine, that contribute to its strong absorption and efficient charge separation. In the conventional solar cell architecture, the layers are stacked from bottom to top as follows: glass substrate, indium tin oxide (ITO) as the transparent anode, PEDOT:PSS as the hole transport layer (HTL), the PM6:Y6 photoactive layer, PNDIT-F3N-Br as the electron transport layer (ETL), and silver (Ag) as the top cathode.
% This configuration collects holes at the bottom and electrons at the top. In contrast, the inverted solar cell structure is organized with the glass and ITO at the bottom, followed by ZnO as the electron transport layer, the PM6:Y6 active layer, MoO$_3$ as the hole transport layer, and Ag as the top electrode. This inverted structure, which collects electrons at the bottom and holes at the top, uses oxide-based interlayers (ZnO and MoO$_3$) known for their stability and charge selectivity but also associated with photocatalytic activity and interfacial redox reactions.
% The figure highlights the diffusion of halogen atoms, particularly fluorine and, in the presence of 1-chloronaphthalene, chlorine, from the photoactive layer to the oxide interlayers in the inverted device. This halogen redistribution is shown via arrows or concentration gradients, suggesting accumulation at the ZnO and MoO$_3$ interfaces, in contrast to the conventional architecture, where the halogen distribution remains uniform within the active layer. This visualization emphasizes the differing interfacial behavior and potential stability issues related to halogen migration in inverted organic solar cells.

\begin{figure}
\includegraphics[width = 0.45\textwidth, height = 0.27\textwidth]{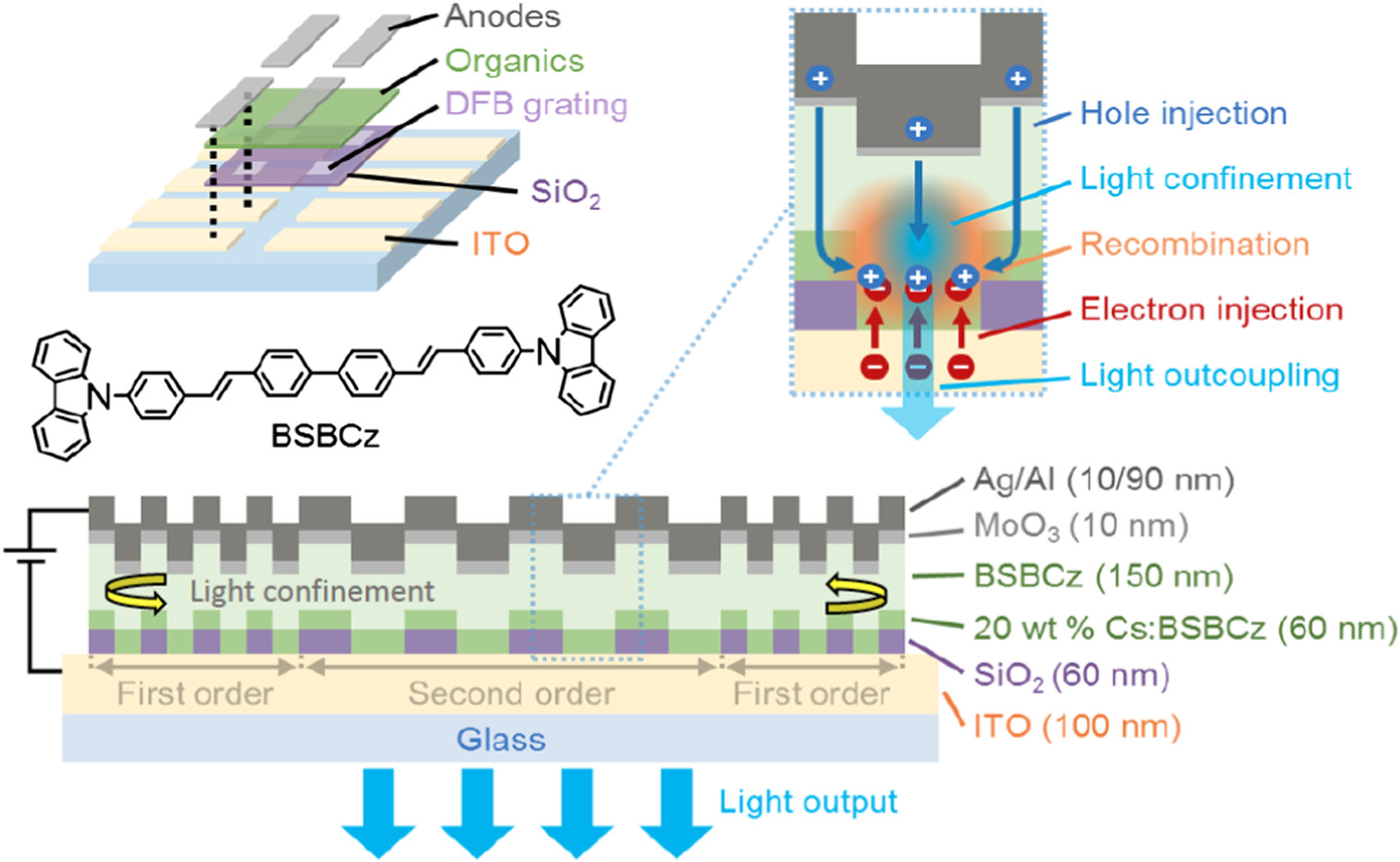}
\caption{an organic semiconductor laser diode (OSLD) device structure based on a single-layer organic semiconductor design, optimized for high-current electrical excitation with low-loss optical feedback via integrated DFB gratings and charge-injection layers that enable effective exciton generation and emission \cite{adachi_organic_led_2020}}
\label{organic led to laser diode, adachi, 2020}
\end{figure}

Figure \ref{organic led to laser diode, adachi, 2020} presents a schematic diagram illustrating the device architecture of a current-injection organic semiconductor laser diode (OSLD). The overall device structure is fundamentally similar to that of a conventional organic light-emitting diode (OLED), with modifications to support electrical pumping and laser oscillation.
The device is structured with an indium tin oxide (ITO) anode at the top and an aluminum (Al) cathode at the bottom. Sandwiched between these electrodes is the organic active layer, which consists of a thin film of the organic semiconductor material BSB-Cz. This BSB-Cz layer functions as the light-emitting medium and is positioned between two distributed feedback (DFB) structures, a primary and a secondary grating, that serve as the optical resonator to achieve laser feedback.
To enhance electrical injection and reduce the contact resistance at the interfaces, specific doping layers are incorporated. On the cathode (bottom) side, the BSB-Cz layer is n-doped with cesium (Cs) to form an ohmic contact with the aluminum cathode. On the anode (top) side, a thin (10 nm) molybdenum trioxide (MoO$_3$) layer is inserted to provide p-type doping and to promote hole injection from the ITO anode into the organic layer.
% The total thickness of the organic active layer is constrained to 210 nm to allow the generation of a high electric field on the order of $10^6$ V/cm, which is required for efficient charge injection and transport in organic amorphous thin-film devices. Unlike conventional OLEDs, which typically use a double heterostructure to confine excitons and improve efficiency, this device employs a single-layer architecture. This structural choice avoids the creation of heterointerfaces that can facilitate non-radiative exciton deactivation processes, especially under the high current densities (up to 1 kA/cm$^2$) needed for laser oscillation.
The BSB-Cz material itself exhibits balanced electron and hole mobilities of approximately $10^{-4}$ cm$^2$/Vs. Given the ohmic contacts at both electrodes, recombination of charge carriers is expected to occur near the center of the BSB-Cz emission layer. Additionally, the external quantum efficiency (EQE) of the device remains nearly constant up to a high current density of around 1 kA/cm$^2$, indicating efficient recombination without significant roll-off.

\subsection{Transition metal dichalcogenides (TMD) solar cells}
\begin{figure}
\includegraphics[width = 0.45\textwidth, height = 0.3\textwidth]{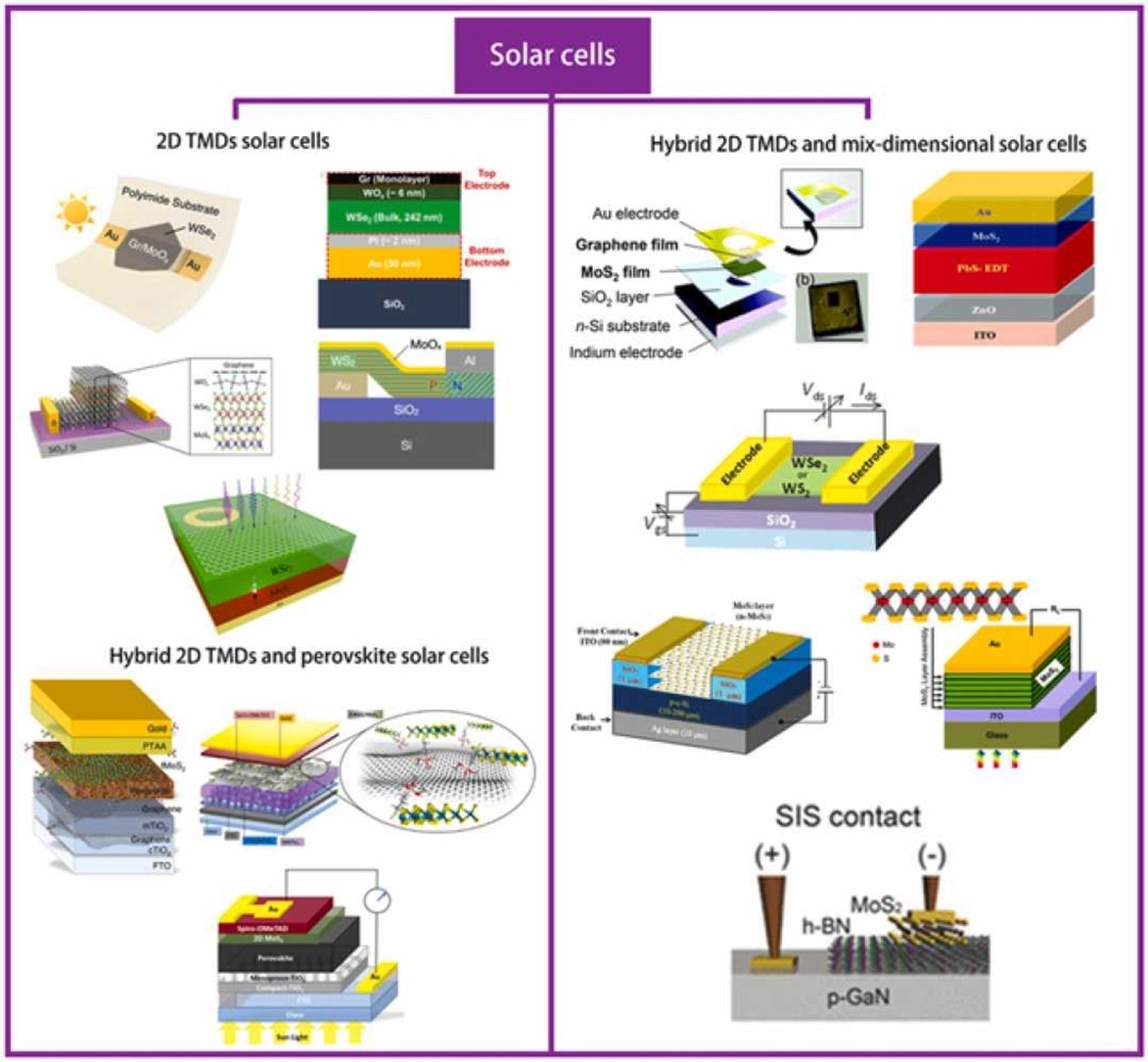}
\caption{2D layered semiconductors for photovoltaic devices, transition metal dichalcogenide and perovskite \cite{aftab_2023}}
\label{dichalcogenide, aftab}
\end{figure}

Two-dimensional TMDs, like MoS$_2$ and WSe$_2$, possess direct bandgaps and exhibit strong excitonic effects, making them suitable for optoelectronic applications \cite{nazif_2021}. Their strong light-matter coupling and exciton dynamics are pivotal for devices such as photodetectors and solar cells \cite{nguyen_2023}. The excitonic properties of TMDs, including bright, dark, and interlayer excitons, play a significant role in their optical response and device performance \cite{friedman_2017}. As shown in figure \ref{dichalcogenide, aftab}, TMDs are suitable materials for advanced photovoltaic (PV) technologies due to their tunable electronic properties, high transparency, and strong light-matter interaction. These materials exhibit high absorption coefficients even at nanoscale thicknesses and are suitable for forming heterostructures without lattice matching. However, challenges such as high exciton binding energies and inefficient charge extraction due to Fermi-level pinning limit their power conversion efficiency (PCE). Strategies including internal electric field engineering, selective passivation contacts, and integration with perovskite solar cells (PSCs) have shown promise in addressing these limitations. The combination of 2D TMDs with perovskites, which offer high mobility and light absorption but suffer from stability issues, has led to enhanced charge separation and improved device longevity, pointing toward viable next-generation solar cell designs.

\begin{figure}
\includegraphics[width = 0.45\textwidth, height = 0.27\textwidth]{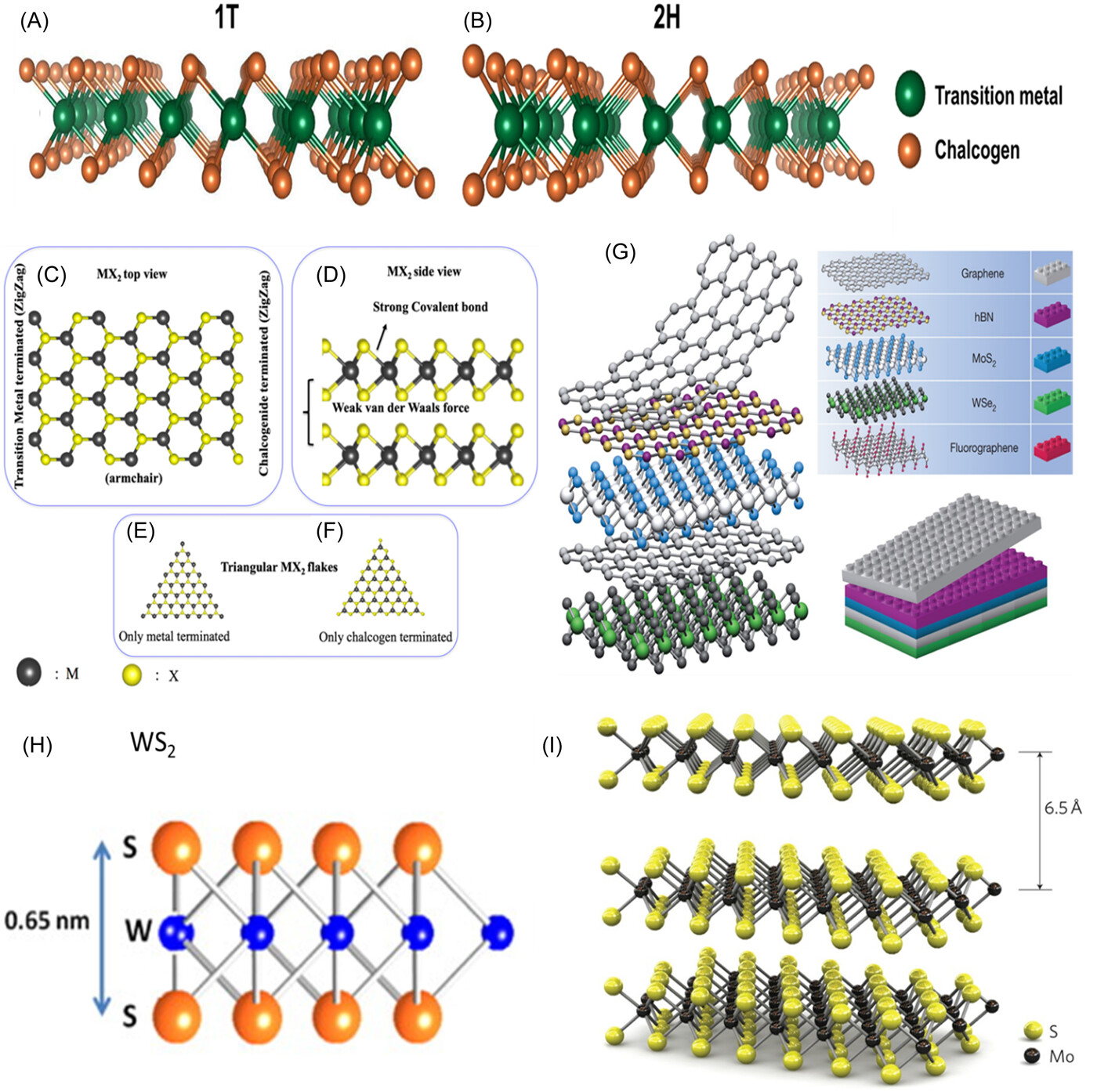}
\caption{transition metal dichalcogenide 1T phase (octahedral structure) and 2H phase (trigonal prismatic structure) \cite{nguyen_2023}}
\label{electrocatalysts, nguyen}
\end{figure}

Figure \ref{electrocatalysts, nguyen} illustrates the structural diversity and material characteristics of transition metal dichalcogenides (TMDs), which follow the general chemical formula MX\textsubscript{2}, where M denotes a transition metal such as Mo, W, Ti, or Nb, and X represents a chalcogen element such as S, Se, or Te. Panels (A) and (B) depict two typical crystal phases of TMDs: the 1T and 2H structures. Both structures consist of monolayers with a sandwich-like X-M-X configuration, in which a hexagonal plane of metal atoms is positioned between two planes of chalcogen atoms. These atoms are covalently bonded, and the metal atoms may adopt either octahedral (1T) or trigonal prismatic (2H) coordination geometries. The electronic properties of TMDs stem from the partially filled d-orbitals of the transition metal atoms, leading to a range of behaviors from semiconducting to superconducting depending on the specific chemical composition. Panels (C) and (D) provide top and side views of a monolayer MX\textsubscript{2}, further emphasizing the structural regularity and symmetry of these materials.
% Panels (E) and (F) show terminations at the transition metal and chalcogen sides, respectively, highlighting edge configurations that play significant roles in catalysis and surface chemistry. Panel (G) demonstrates heterostructured stacking of different TMD monolayers, which enables the engineering of diverse material properties such as photovoltaic response, thermal conductivity, elasticity, and catalytic activity. The intrinsic characteristics of TMDs, including direct bandgaps in group-VI monolayers (e.g., MoS\textsubscript{2}, WS\textsubscript{2}, MoSe\textsubscript{2}, WSe\textsubscript{2}), strong spin-orbit coupling, and van der Waals-layered structures, make them suitable for various electrochemical applications like hydrogen evolution reactions (HER), batteries, and supercapacitors. Finally, panels (H) and (I) provide detailed views of the layered architectures of two prototypical TMDs: tungsten disulfide (WS\textsubscript{2}) and molybdenum disulfide (MoS\textsubscript{2}), which further exemplify the tunability and anisotropic properties of these layered materials.
% \begin{figure}
% \includegraphics[width = 0.45\textwidth, height = 0.37\textwidth]{image/antisolventsolarcell,ying2025.png}
% \caption{\cite{ying_shang_2025}}
% \end{figure}
% \begin{figure}
% \includegraphics[width = 0.45\textwidth, height = 0.3\textwidth]{image/flexiblequantumdot,longhu2021.png}
% \caption{\cite{long_hu_quantum_dot_2021}}
% \end{figure}

\subsection{Cadmium telluride (CdTe) solar cells}

CdTe is a prominent material in thin-film solar cells, known for its near-optimal bandgap ($\sim$1.5 eV) and high absorption coefficient. Recent designs incorporating advanced contact materials and buffer layers have achieved simulated efficiencies up to 31.82\%, with nearly 100\% quantum efficiency at visible wavelengths. These advancements underscore the material's potential for high-efficiency photovoltaic applications.

\subsection{Silicon solar cells}

Silicon remains the most widely used material in photovoltaic technology. While approaching its theoretical efficiency limits, innovations such as tandem configurations with perovskite layers and bifacial designs have been employed to enhance performance. For example, perovskite/silicon tandem cells utilizing luminescent coupling have shown that more than 50\% of excess electron-hole pairs in the perovskite top cell can be utilized by the silicon bottom cell, indicating a significant potential for efficiency improvements.

\subsection{Machine learning (ML) for solar cell materials}
Machine learning (ML) is becoming increasingly pivotal in advancing photovoltaic materials discovery and device optimization \cite{anand_topo_ml_2022}. In the context of perovskite and organic solar cells, ML accelerates the screening of chemical and structural candidates by learning from large datasets derived from density functional theory (DFT) calculations and experimental measurements \cite{tao_2021}, \cite{fiedler_2022}. By integrating ML with high-throughput DFT workflows, researchers have successfully identified stable, efficient compositions within the vast chemical space of perovskites, such as novel inorganic double perovskites with optimal bandgaps and enhanced stability \cite{chen_peroskite_2023}. This data-driven approach enables rapid material discovery without the need for exhaustive trial-and-error experimentation \cite{laakso_2022}.

\subsubsection{Density functional theory (DFT)}

Density functional theory is used to analyze the electronic structure of many-body systems using functionals of the spatially dependent electron density. DFT is built upon the Hohenberg-Kohn theorems and the Kohn-Sham approach. The first Hohenberg-Kohn theorem states that the ground-state energy of a many-electron system is a unique functional of the electron density \( \rho(\mathbf{r}) \). The second theorem provides a variational principle
\[
E_0 = \min_{\rho} \left\{ F[\rho] + \int V_{\text{ext}}(\mathbf{r}) \rho(\mathbf{r}) \, d\mathbf{r} \right\}
\]
where \( F[\rho] \) is a universal functional of the density, composed of the kinetic energy and electron-electron interaction energy. In the Kohn-Sham approach, the interacting system is mapped onto a non-interacting system with the same density, leading to the Kohn-Sham equations
\begin{align*}
&\left[ -\frac{\hbar^2}{2m} \nabla^2 + V_{\text{eff}}(\mathbf{r}) \right] \psi_i(\mathbf{r}) = \epsilon_i \psi_i(\mathbf{r})\\
&\rho(\mathbf{r}) = \sum_{i=1}^{N} |\psi_i(\mathbf{r})|^2\\
&V_{\text{eff}}(\mathbf{r}) = V_{\text{ext}}(\mathbf{r}) + \int \frac{e^2 \rho(\mathbf{r}')}{|\mathbf{r} - \mathbf{r}'|} \, d\mathbf{r}' + V_{\text{xc}}[\rho](\mathbf{r})
\end{align*}
where \( V_{\text{xc}}[\rho] \) is the exchange-correlation potential, usually approximated using functionals like local density approximation
(LDA), generalized gradient approximation (GGA), or hybrid functionals (e.g., PBE0, HSE06).

\subsubsection{Density functional theory (DFT) in machine learning for solar cell materials}

Machine learning (ML) models are trained on DFT-computed quantities such as formation energy \( E_f \), bandgap \( E_g \), dielectric constants, effective masses, and charge densities. The ML workflow typically involves a descriptor \( \mathbf{x} \) derived from atomic structure and electronic properties, and a target property \( y \) predicted via a regression model \( y = f(\mathbf{x}) \). The learning objective is
\[
\min_{f \in \mathcal{F}} \sum_{i=1}^N \left( y_i - f(\mathbf{x}_i) \right)^2 + \lambda \Omega(f)
\]
where \( \Omega(f) \) is a regularization term and \( \mathcal{F} \) is a function space determined by the ML model (e.g., kernel ridge regression, random forest, neural networks).

In perovskite and organic photovoltaics, DFT-ML pipelines are used for absorber layer screening by predicting the bandgap \( E_g \) in the optimal range (1.1-1.6 eV), stability (formation enthalpy), and defect tolerance. The data set includes chemical formulas, structure fingerprints, and DFT-calculated properties.
Additive screening involves modeling how small molecules alter the electronic environment of the active layer, affecting charge recombination and phase stability. ML models trained on DFT-predicted charge densities and dipole moments are used to classify or regress performance-enhancing behavior.
In dichalcogenides and other layered semiconductors, DFT is used to calculate charge carrier effective masses \( m^* \), exciton binding energies \( E_b \), and anisotropic mobility tensors. ML models then correlate these with experimental device metrics.

Electron transport simulations use quantities like the mobility tensor \( \mu_{ij} \), derived from deformation potential theory or Boltzmann transport equations using DFT data
\[
\mu = \frac{2e\hbar^3 C}{3k_B T |m^*|^2 E_1^2}
\]
\( C \) is the elastic modulus and \( E_1 \) the deformation potential, both computable from DFT.

DFT combined with ML is also used to
predict ion migration barriers using nudged elastic band (NEB) calculations,
screen interface dipoles and band alignments at heterojunctions,
evaluate thermodynamic stability across phase diagrams,
design hole/electron transport materials by predicting HOMO/LUMO levels,
and reorganization energies from Koopmans-compliant DFT.

\subsection{Machine learning for perovskite solar cell}
\begin{figure}
\includegraphics[width = 0.45\textwidth, height = 0.27\textwidth]{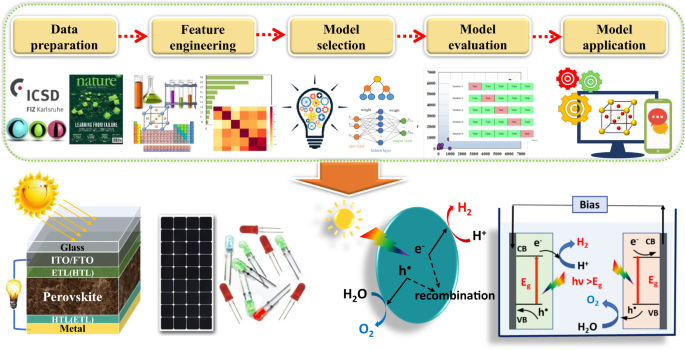}
\caption{ML workflow in perovskite to screen materials with target properties \cite{tao_2021}}
\label{qiuling tao, ml perovskite, 2021}
\end{figure}
ML can be used to enhance additive screening by predicting the impact of thousands of potential organic molecules on perovskite film formation and stability \cite{zhang_2022_add}, a perovskite ML workflow is shown in figure \ref{qiuling tao, ml perovskite, 2021}. For example, ML-guided screening of over 250,000 organic additives led to the discovery of promising passivation agents that significantly boost power conversion efficiencies (PCEs) \cite{xia_2024}, some achieving values over 25\% \cite{zhi_2023_screen}. Neural networks trained on photophysical properties such as transmission and photoluminescence spectra have been used to predict key device parameters including open-circuit voltage, short-circuit current, and fill factor \cite{zhang_halide_2025}. ML models have been employed to design nanostructured optical architectures, like pyramid textures, enhancing light absorption and boosting PCE to 28.4\% in simulations \cite{li_arch_2024}. ML algorithms analyze experimental data to optimize fabrication parameters, improving device performance and stability \cite{karade_2023}. High-throughput DFT calculations and ML can be used to predict electronic properties of TMD/perovskite heterostructures, identifying combinations with favorable band alignments for efficient solar cells. ML models, such as crystal graph convolutional neural networks, predict bandgaps and alignments in thousands of TMD/perovskite combinations, guiding experimental efforts \cite{congsheng_xu_2024}. These predictions facilitate real-time optimization and performance estimation of experimental devices, allowing researchers to focus on the most promising formulations and architectures.

\subsection{Machine learning for organic solar cell (OSC)}
In organic solar cells (OSC), ML models trained on molecular descriptors of donor-acceptor systems have demonstrated remarkable accuracy in predicting power conversion efficiency (PCE) \cite{ahmed_2025}, energy levels, and absorption spectra \cite{greenstein_2023}. An example is the use of ML to predict PCEs of various polymer:non-fullerene acceptor combinations, leading to the fabrication of devices with efficiencies up to 15.23\% \cite{suthar_2023}. These models, including deep learning and ensemble approaches \cite{wang_ensemble_2023}, allow for the efficient virtual screening of materials, often revealing candidates with high predicted efficiencies that had not yet been synthesized \cite{zhang_cai_organic_2023}. ML also enables reverse design strategies, in which target optoelectronic properties guide the molecular structure search, thus inverting the traditional materials development workflow.

ML is instrumental in optimizing device architecture and fabrication conditions. Supervised learning algorithms such as support vector machines (SVMs), random forests (RFs), and convolutional neural networks (CNNs) are applied to experimental datasets to predict optimal layer thicknesses, interface compositions, and annealing conditions \cite{cetinkaya_2025}. These algorithms are also employed to predict key performance metrics (PCE, open-circuit voltage) based on molecular descriptors and device parameters \cite{das_2024}. These models analyze the relationship between molecular structure and device performance, guiding the design of new devices \cite{mahmood_2021}. This helps streamline fabrication workflows and improve device reproducibility and stability. Optical design has also benefited from ML-guided architecture optimization \cite{meftahi_2020}, such as the creation of nanostructured light-trapping textures that enhance absorption and push simulated PCEs beyond 28\%.

The combination of ML with computational methods such as DFT and drift-diffusion simulations offers a powerful framework for predicting electronic properties, carrier mobilities, exciton dissociation rates, and energy-level alignments. These predictions guide the experimental selection of material combinations and interfacial engineering strategies in multilayered devices. Recent work has demonstrated the use of crystal graph neural networks to model the electronic structure of thousands of TMD/perovskite heterostructures, identifying optimal configurations for improved charge transfer and minimized recombination losses.

\begin{figure}
\includegraphics[width = 0.45\textwidth, height = 0.3\textwidth]{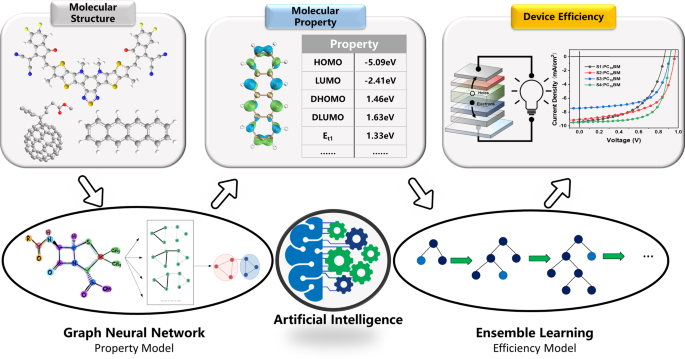}
\caption{molecules are converted to graphs inputs in graph neural network property model, power conversion efficiency predicted by efficiency model \cite{wang_ensemble_2023}}
\label{hongshuai wang, efficient screening framework, 2023}
\end{figure}

Figure \ref{hongshuai wang, efficient screening framework, 2023} illustrates the workflow designed for efficient prediction of power conversion efficiencies (PCEs) in organic solar cells (OSCs). The workflow comprises two interconnected modeling components: the Property Model and the Efficiency Model. The Property Model is based on Graph Neural Networks (GNNs) and is responsible for predicting molecular properties from molecular structures. To do this, molecular structures are first converted into graph representations that serve as inputs to the GNN. The GNN then outputs relevant physicochemical properties of the molecules, which serve as features for the subsequent modeling stage. The second component, the Efficiency Model, is built using an ensemble learning method, specifically the Light Gradient Boosting Machine (LightGBM). This model takes as input the molecular properties predicted by the Property Model and outputs the predicted PCEs of the corresponding OSC devices. In this way, the Efficiency Model maps molecular-level descriptors to macroscopic device-level performance metrics. To construct the models, two databases were used. The first is a curated dataset containing both experimental device efficiencies and physicochemical properties obtained via density functional theory (DFT) calculations. This dataset was used to train the Efficiency Model, thereby capturing the relationship between molecular properties and device efficiency. The second dataset, used to train the Property Model, comprises a large number of molecular structures along with their corresponding DFT-derived properties. This allows the Property Model to bypass the computational cost of DFT by accurately predicting molecular properties directly from structure. By integrating these two models, the workflow achieves both speed and accuracy in predicting OSC performance. The Property Model enables fast estimation of molecular properties without expensive computations, while the Efficiency Model ensures accurate predictions of PCEs based on those properties. This combined framework allows for a direct and streamlined approach to predicting OSC efficiencies from molecular structures alone, facilitating rapid screening and optimization of new materials.

\section{Conclusion}

% Quantum optoelectronics holds promise for revolutionizing various technological domains. In quantum communication, it enables the development of secure communication channels using quantum key distribution and entangled photon sources. In quantum computing, it facilitates the implementation of quantum bits (qubits) using photonic systems for scalable quantum processors. In precision metrology, it enhances measurement sensitivity in applications like gravitational wave detection and atomic clocks. In quantum networks, it establishes interconnected quantum systems for distributed computing and information sharing.

% The integration of quantum optoelectronic principles into semiconductor solar cells holds the promise of surpassing current efficiency limits and enabling new functionalities. By leveraging cavity effects, exciton dynamics, and advanced material properties, we can develop next-generation photovoltaic devices with superior performance and novel capabilities. Continued interdisciplinary research combining materials science, quantum optics, and device engineering will be crucial in realizing these advancements. Continued research focuses on integrating quantum optoelectronic components into scalable, chip-based platforms, improving coherence times, and developing robust fabrication techniques for practical deployment.

This paper has demonstrated that the future solar cell performance can be improved by a combination of quantum optics and semiconductor photonics. By incorporating resonant optical cavities, strong light-matter coupling, and quantum photonic control into photovoltaic materials, particularly perovskites, organics, transition metal dichalcogenides (TMD), and hybrid structures, significant gains in optical absorption, carrier lifetime, and voltage can be achieved. Emerging experimental techniques, including frequency comb spectroscopy and laser interferometry, provide novel tools for both device diagnostics and fundamental studies.

In the future, artificial intelligence are going to provide significant guidance to the development of solar cells. Machine learning frameworks trained on DFT-calculated properties allow for rapid screening of materials and interface designs, reducing reliance on trial-and-error experimentation. These findings collectively suggest that a quantum-optical approach to solar energy research not only offers practical efficiency improvements but also establishes a platform for deeper integration of semiconductor device physics, nanophotonics, and computational materials science in renewable energy technologies.

The development of next generation high efficiency, lightweight, and flexible optoelectronic devices, batteries, and solar cells, continues to advance through innovations in materials science, nanofabrication, and computational modeling. Progress in device fabrication methods, such as photolithographic patterning of electrodes, spin-coating and atomic-layer deposition of transport-layer materials, and advanced characterization of device performance, efficiency, and long term stability has enabled more refined control over architecture and functionality. Computational approaches, including density-functional theory (DFT) and drift diffusion simulations, are increasingly used to predict optimal energy-level alignments, carrier mobilities, and exciton dissociation efficiencies, thereby guiding experimental efforts in material selection and interface design. Research in this field increasingly integrates solution processed organic semiconductors with high crystallinity inorganic films in tandem architectures, aiming to achieve high power conversion efficiencies and extended operational lifetimes under continuous illumination. These strategies have the potential to reduce manufacturing costs and accelerate the deployment of versatile photovoltaic technologies, from lightweight and flexible modules to building integrated systems. The convergence of machine learning driven materials discovery, performance prediction, and process optimization with hands on nanofabrication and advanced characterization is shaping a new paradigm. This integrative approach supports the development of photovoltaic devices that are not only highly efficient and stable but also scalable, cost-effective, and environmentally sustainable.
\section{Declaration of competing interests}
The authors declare there is no competing interest.
\bibliographystyle{elsarticle-num}
\bibliography{ref}
\end{document}